\documentclass[aps,prl,twocolumn,showkeys,amsmath,amssymb,longbibliography,floatfix,superscriptaddress]{revtex4-1}

\usepackage[utf8]{inputenc}
\usepackage[T2A]{fontenc}
\usepackage{appendix}
\usepackage{amsmath,amssymb}
\usepackage{physics}

\usepackage{graphicx}

\usepackage{xcolor}
\usepackage[colorlinks=true, urlcolor=blue, linkcolor=red, citecolor=blue]{hyperref}

\usepackage[main=english,russian]{babel}

\usepackage{float}
\usepackage{booktabs}
\usepackage[normalem]{ulem}
\usepackage{cancel,soul}
\usepackage{tikz}

\begin{document}

\title{Conformal Data for the O(3) Wilson-Fisher Conformal Field Theory from Fuzzy Sphere Realization of the Quantum Rotor Model}

\author{Arjun Dey}
\affiliation{
Laboratory for Theoretical and Computational Physics,
PSI Center for Scientific Computing, Theory and Data,
5232 Villigen PSI, Switzerland}
\affiliation{
Institute of Physics,
\'{E}cole Polytechnique F\'{e}d\'{e}rale de Lausanne (EPFL),
1015 Lausanne, Switzerland}

\author{Loic Herviou}
\affiliation{Univ. Grenoble Alpes, CNRS, LPMMC, 38000 Grenoble, France}

\author{Christopher Mudry}
\affiliation{
Laboratory for Theoretical and Computational Physics,
PSI Center for Scientific Computing, Theory and Data,
5232 Villigen PSI, Switzerland}
\affiliation{
Institute of Physics,
\'{E}cole Polytechnique F\'{e}d\'{e}rale de Lausanne (EPFL),
1015 Lausanne, Switzerland}

\author{Andreas Martin L\"auchli} %{Andreas Martin L\"auchli Herzig}
\affiliation{
Laboratory for Theoretical and Computational Physics,
PSI Center for Scientific Computing, Theory and Data,
5232 Villigen PSI, Switzerland}
\affiliation{
Institute of Physics,
\'{E}cole Polytechnique F\'{e}d\'{e}rale de Lausanne (EPFL),
1015 Lausanne, Switzerland}

\begin{abstract}
We present a model for strongly interacting fermions with internal
O(3) symmetry on the fuzzy sphere
that (i) preserves the rotational symmetry of the fuzzy sphere
and (ii) undergoes a quantum phase transition in the
(2+1)-dimensional O(3) Wilson-Fisher universality class.
Using exact diagonalization (ED)
and density matrix renormalization group (DMRG),
we locate the quantum critical point via conformal
perturbation theory and obtain scaling dimensions from finite-size
spectra. 
We identify 24 primary operators and determine some of their
operator product expansion coefficients through first-order conformal
perturbation theory. The results are benchmarked
against conformal bootstrap
and large quantum-number expansions
and reveal a weakly irrelevant operator
that plays a role in dimerized antiferromagnets. Our work
provides a general framework for quantitatively accessing conformal data for $O(N)$ Wilson-Fisher conformal field theories (CFTs).
\end{abstract}
    
\maketitle

\textit{Introduction.---} Conformal field theories (CFTs)
in $(d+1)$-dimensional spacetime [$(d+1)$D] 
are fully specified by their conformal data, scaling dimensions, and
operator product expansion (OPE) coefficients, which govern universal
critical behavior, and some measurable
responses~\cite{PismaZhETF,Di_Francesco_Mathieu_Senechal_1997,Rychkov_2017,Cardy_1996,goldenfeld2018lectures,Witczak_Krempa_2012,Sachdev_2011}.
While (1+1)D CFTs~\cite{Di_Francesco_Mathieu_Senechal_1997, Belavin_Polyakov_Zamolodchikov_1984} enjoy remarkable theoretical completeness through
Virasoro algebra~\cite{Virasoro_1970} and integrability~\cite{Takahashi_1999}, the landscape of (2+1)D CFTs
remains largely unexplored, presenting rich opportunities for
discovering new universal physics and testing fundamental principles
of quantum field theory.

Standard approaches have limitations when it comes to accessing
detailed information about operator spectra and OPE data of (2+1)D
CFTs, especially when they are non-Abelian. Large-scale Monte Carlo simulations~\cite{Hasenbusch_2022} pin down critical exponents but have trouble resolving subleading operators or in extracting  OPE data. Conformal bootstrap~\cite{Vichi_Slava_2011, Kos_Poland_Simmons-Duffin_Vichi_2016} studies of non-Abelian theories provide only a limited set of low-lying operators~\cite{O3Bootstrap_Shai}. Perturbative analytics rely on resummations that lose accuracy for spinning operators~\cite{Derkachov_Manashov_1997_epsilon,LANG1993573_1_N, Henriksson:2025hwi}. Fuzzy-sphere regularization~\cite{Zhu_Han_Huffman_Hofmann_He_2023, Fuzzy_OPE,
Fuzzy_4pt_correlators, Fuzzy_Defects, Fuzzy_g_function,
Fuzzy_Entropic_F_function, Fuzzy_Impurities, Fuzzy_Surface_CFTs,
Fuzzy_Generators, Fuzzy_SO5, FuzzyRealScalarHe,
FuzzyRealScalarTaylor} now provides a fully symmetric Hamiltonian probe of spectra and correlators and even some OPE data.
A prior fuzzy sphere study of the
(2+1)D CFT with internal O(3) symmetry confirmed feasibility but identified only a limited number
of operators~\cite{Conformal_Fuzzy_Wilson_Fisher_Content}.

Here, we introduce a
fuzzy sphere realization (FZR)
of a truncated quantum rotor model (TQRM)
~\cite{Sachdev_2011,Zhu_Han_Huffman_Hofmann_He_2023}
that retains the full internal O(3) and spatial SO(3) symmetry
of the (2+1)-dimensional O(3) Wilson-Fisher (WF) universality class
and is tractable to ED and DMRG. Building on the well‐known
correspondence between O(3) quantum rotors and O(3)
quantum criticality~\cite{Sachdev_2011},
the idea is to use a conformal map to identify a compactification
of two-dimensional space with the unit 2-sphere $\mathbb{S}^2$ and
to leverage the state-operator correspondence to extract conformal data.
This is done by relating the finite-size gaps of the FZR of the TQRM
to the scaling dimensions of the CFT operators.
We identify 24
low-lying primary operators, with their scaling dimensions summarized
in Tables~\ref{tab:bs_data_table}
and~\ref{tab: primaries table} (Appendix
\ref{sec: appendix Primary operators}).
Selected OPE coefficients are also determined. The full set
of scaling dimensions and OPE values can be found in the Supplementary
Material~\cite{SuppMat}.  We benchmark our results with bootstrap
results~\cite{O3Bootstrap_Shai} and large quantum-number expansion
predictions~\cite{Cuomo2020,Cuomo2021}, wherever they are available.

\textit{Quantum rotor model.---} A rigid rotor, a particle of mass
\(\mu\) constrained to move on the surface of a sphere of radius
\(R\), serves as the prototypical quantum model of rotational motion.
Its energy is purely kinetic, arising from quantized angular momenta
with moment of inertia
\(I=\mu\,R^2>0\).
On a square lattice, an array of such
rotors with nearest-neighbor ferromagnetic Heisenberg coupling $J>0$
realizes an O(3) magnetic transition when
the dimensionless coupling $I\,J$
is tuned to quantum criticality~\cite{Sachdev_2011}.

The full two-dimensional interacting lattice Hamiltonian for
O(3) quantum rotors reads
$\widehat{H}_{\infty}:=
\widehat{H}_{\mathrm{kin}}
+
\widehat{H}_{\mathrm{int}},
$
\begin{subequations}
\begin{equation}
\begin{split}
&
\widehat{H}_{\mathrm{kin}}:=
\frac{1}{2I}\sum_{i=1}^{N}
\widehat{\mathbf{L}}_i^2, 
\qquad
\widehat{H}_{\mathrm{int}}:=
-
J
\sum_{\langle ij\rangle}
\hat{\mathbf{n}}_i
\cdot
\hat{\mathbf{n}}_j,
\end{split}
\end{equation}
where the pair of operator-valued 3-component vectors
$(\widehat{\mathbf{L}}_i,\hat{\mathbf{n}}_i)$
obey the equal-time O(3) rotor algebra
\begin{equation}
\begin{split}
&
[\widehat{L}_i^a,\widehat{L}_j^b]:=
\mathrm{i}\hbar\,\epsilon^{abc}\,\widehat{L}_i^c\,\delta_{ij}, \quad [\hat{n}_i^a,\hat{n}_j^b]:=0,
\\
&
[\widehat{L}_i^a,\hat n_j^b]:=
\mathrm{i}\hbar\,\epsilon^{abc}\,\hat n_i^c\,\delta_{ij},
\quad
a,b=
1,2,3\equiv
x,y,z,
\end{split}
\end{equation}
with the local constraints
\begin{equation}
\widehat{\mathbf{L}}_i^2=
\hbar^{2}\,l(l+1)\,
\widehat{\mathbb{I}},
\qquad
\hat{\mathbf{n}}_i^2:=
\widehat{\mathbb{I}}.
\end{equation}
\end{subequations}
Here, $l\in\mathbb{Z}_{\geq0}$ fixes a global representation
of each quantum rotor, $\langle ij\rangle$ are nearest-neighbors sites
on the square lattice with $N$ the number of sites.

In the limit $I\,J=\infty$, $\widehat{H}_{\infty}$ simplifies to the
classical O(3) ferromagnetic Heisenberg model
\begin{equation}
\widehat{H}_{\mathrm{int}}=
-
J
\sum_{\langle ij\rangle}
\mathbf{n}_i
\cdot
\mathbf{n}_j,
\qquad
\mathbf{n}_i^2=1,
\end{equation}
with the ferromagnetic ground state
$|\mathbf{n}_1,\ldots,\mathbf{n}_N\rangle=
|\mathbf{n},\ldots,\mathbf{n}\rangle$
whereby
$\hat{\mathbf{n}}_i\,
|\mathbf{n}_1,\ldots,\mathbf{n}_N\rangle=
\mathbf{n}_i\,
|\mathbf{n}_1,\ldots,\mathbf{n}_N\rangle$.
In the limit $I\,J=0$, $\widehat{H}_{\infty}$ simplifies to
\begin{align}
\widehat{H}_{\mathrm{kin}}=
\sum_{i=1}^{N}
\sum_{l=0}^{\infty}
\sum_{m_i=-l}^{+l}
\frac{\hbar^2\,l(l+1)}{2I} \ket{l,m_i} \bra{l, m_i}
\label{eq:diagonal Hkin}
\end{align}
with \(\ket{l,m_i}\)
the eigenstate of $\widehat{\mathbf{L}}_i^2$ and $\widehat{\mathbf{L}}_i^z$
with eigenvalues $\hbar^{2}\,l(l+1)$ and $\hbar\,m_i$, respectively.
The ground state of $\widehat{H}_{\mathrm{kin}}$
is the singlet state with $l=0$. It is believed that,
in the thermodynamic limit $N\uparrow\infty$, 
there exists an unstable quantum critical point when $I\,J\approx 1$
that separates the long-range ferromagnetically ordered phase
controlled by the fixed point Hamiltonian 
$\widehat{H}_{\mathrm{int}}$
from the paramagnetic phase 
controlled by the fixed point Hamiltonian 
$\widehat{H}_{\mathrm{kin}}$. This quantum critical point is
believed to belong to the (2+1)D O(3) WF universality class.

Numerical simulations of a lattice Hamiltonian requires a Hilbert space
of finite dimensionality. We therefore truncate each quantum rotor
on the right-hand side of Eq.\ (\ref{eq:diagonal Hkin})
at a maximum angular momenta \(l_{\mathrm{max}} = 1 \),
retaining all states \(\ket{l,m}\) with \(0\le l\le l_{\mathrm{max}}\) and
\(-l\le m\le l\). 
This procedure, by which $\widehat{H}_{\infty}$ becomes
$\widehat{H}_{l_{\mathrm{max}}}$,
preserves the full O(3) symmetry,
since each $l$ multiplet is kept intact~\cite{SuppMat}.
Moreover, this procedure is not expected to change the topology of
the phase diagram at vanishing temperature.

\textit{Fuzzy sphere~many-body~electronic~Hamiltonian.---}
To access conformal data, we map the quantum rotor model to a system
of $N$ fermions with internal degrees of freedom, moving on a
sphere~\cite{Haldane_1983_FQHE} and projected to the lowest Landau
level (LLL). In this limit, the noncommuting coordinate operators
generate the fuzzy sphere algebra~\cite{JMadore1992},
thereby providing a natural
short-distance cutoff
(the magnetic length) that preserves rotational symmetry.

We focus on the case where each fermion has $4$ flavors and the
system resides at quarter filling. At quarter filling, strong
interactions enforce (on average) single occupancy per LLL orbital,
analogous to the large-$U$ limit of the Hubbard model.

The microscopic fuzzy sphere Hamiltonian
that we shall study numerically is defined to be
\begin{equation}
\widehat{H}_{\mathrm{fzs}}:=
\widehat{P}_{\mathrm{LLL}}
\left(
u\,
\widehat{H}_{\mathrm{Hub}}
-
v\,
\widehat{H}_{\mathrm{Heis}}
+
h\,
\widehat{H}_{\mathrm{trans}}
\right)
\widehat{P}_{\mathrm{LLL}}.
\label{eq:def H fs}
\end{equation}
Here, $\widehat{P}_{\mathrm{LLL}}$ denotes the projector to the LLL.
The dimensionless couplings
$\delta u\equiv u-u_{\mathrm{c}}$,
$\delta v\equiv v-v_{\mathrm{c}}$,
$\delta h\equiv h-h_{\mathrm{c}}\in\mathbb{R}$
measure the deviations away from a putatitive
quantum critical point $u_{\mathrm{c}},v_{\mathrm{c}},h_{\mathrm{c}}$
in the (2+1)D O(3) WF universality class.
Hamiltonian
$\widehat{H}_{\mathrm{Hub}}$
is a short-range two-body repulsive U(4)-flavor-symmetric
density-density interaction between the fermions
that implements the constraint that every LLL are singly occupied
in the limit $u\uparrow\infty$, $v,h$ fixed.
Hamiltonian
$\widehat{H}_{\mathrm{Heis}}$
is a short-range two-body Heisenberg-like interaction that selects
a ground-state that breaks spontaneously the internal SO(3) symmetry
down to the subgroup SO(2) in the limit $v\uparrow\infty$,
$u,h=0$ fixed. 
Hamiltonian
$\widehat{H}_{\mathrm{trans}}$
is a one-body term that breaks the
U(4)-flavor-symmetry down to the subgroup U(1)$\times$U(3)
that selects a non-degenerate gapped singlet ground state
in the limit $|h|\uparrow\infty$, $u,v$ fixed. The information about the quantum rotor is encoded inside $\widehat{H}_{\mathrm{Heis}}$ and $\widehat{H}_{\mathrm{trans}}$.
The explicit representation of
$\widehat{H}_{\mathrm{fzs}}$
can be found in the Supplementary Material~\cite{SuppMat}.

\begin{table}[t!]
\centering
\caption{
Summary of selected primary operators.
$S$ labels the internal SO(3) spin representation,
$L$ the SO(3) angular momenta on the fuzzy sphere,
and $\pm$ the $\mathbb{Z}_2$ parity (improper part) of the
internal O(3) symmetry, so that states are labeled by $S^{\pm}$ and $L$.
The index I orders energy eigenstates within each symmetry sector.
Scaling dimensions are extracted via ED
and DMRG,
and are compared to conformal bootstrap (CB); asterisks denote
exact values, and parentheses indicate system size. The two bold-faced
entries in the $\Delta$ (CB) column are used as input.
        }
\begin{tabular}{|ccc|c|c|c|c|}
\hline
$S^{\pm}$ & $L$ & $I$ &$\mathrm{o}$ & $\Delta$(ED) & $\Delta$(DMRG) &
$\Delta$(CB)
\\
\hline
\hline
$1^{-}$ & 0 & 1 & $\sigma$ & 0.51893 (12) & 0.51893 (28) &
\textbf{0.518936}~\cite{O3Bootstrap_Shai}
\\
$0^{+}$ & 0 & 2 & $\varepsilon$ & 1.56189 (12) & 1.61781 (26) &
\textbf{1.59488}~\cite{O3Bootstrap_Shai}
\\
$0^{+}$ & 0 & 4 & $\varepsilon'$ & 3.77279 (12) & 3.80177 (14) &
3.76680~\cite{HENRIKSSON20231}
\\
$0^{+}$ & 2 & 1 & $T_{\mu \nu}$ & 2.97749 (12) & 3.01062 (20) &
$3^{*}$
\\
$1^{+}$ & 1 & 1 &$j_{\mu}$ & 1.88253 (12) & 1.97204 (26) &
$2^{*}$
\\
$2^{+}$ & 0 & 1 & $t_{(2)}$ & 1.25407 (12) & 1.23767 (26) &
1.20954~\cite{O3Bootstrap_Shai}
\\
$4^{+}$ & 0 & 1 & $t_{(4)}$ & 3.26858 (11) & 3.16459 (26) &
2.99056~\cite{O3Bootstrap_Shai}
\\
\hline
\end{tabular}

\label{tab:bs_data_table}
\end{table}

\begin{figure}[t!]
\centering
\includegraphics[width=\columnwidth]{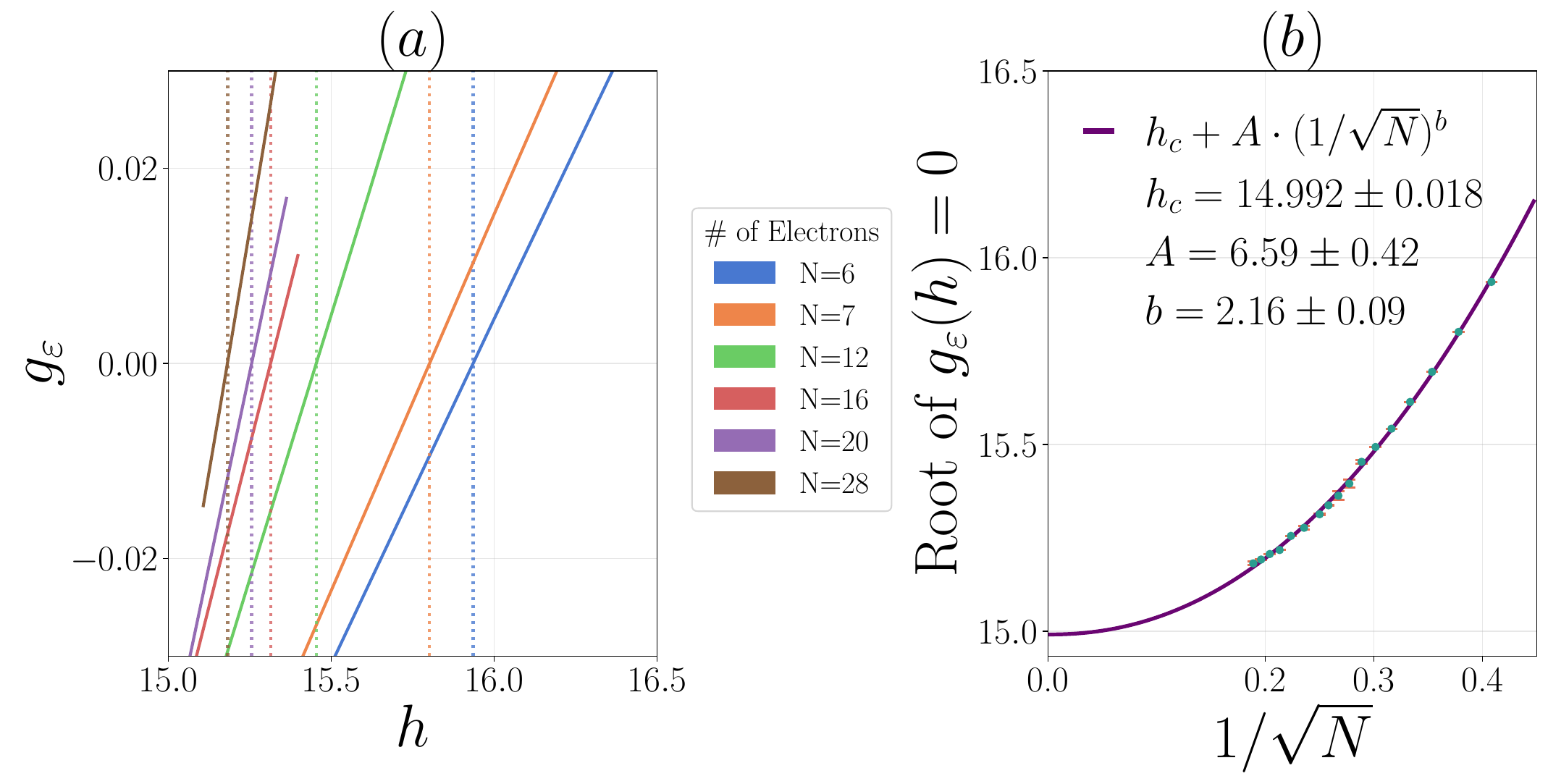}
\caption{
Panel (a) shows the dependence on $h$ of the
relevant coupling $g_{\varepsilon}(h)$ for a given
$R\propto\sqrt{N}$. The critical point $h_{\mathrm{c}}$
is the root to  $g_{\varepsilon}(h)=0$
after the thermodynamic limit $N\uparrow\infty$ has
been taken.
Panel (b) shows the dependence of the root to
$g_{\varepsilon}(h)=0$
as a function of $1/\sqrt{N}\propto 1/R$. 
Fitting this dependence with an appropriate power law
and using the first irrelevant anomalous dimension
$0.7688$~\cite{HENRIKSSON20231}, we extract the critical value
$h_{\mathrm{c}}=14.992\pm0.018$ and the critical exponent
$1/\nu=1.3908\pm0.0852$, which is
in close agreement with the known value
$1.406$~\cite{one_by_nu_hasenbuch}.
        }
\label{fig:critical-points}
\end{figure}

\textit{Symmetries.---}
Hamiltonian (\ref{eq:def H fs})
exhibits three symmetries: (a) spatial rotational symmetry
SO(3), arising from conservation of angular momenta
  with the
quantum number $L$; (b) internal SO(3) spin symmetry
with the quantum number $S$; (c) and a $\mathbb{Z}_2$ symmetry
corresponding to the conservation of the parity of the number of
electrons occupying $p$ orbitals in internal spin space. The quantum
number for the $\mathrm{SO}(3)\times\mathbb{Z}_2\cong\mathrm{O}(3)$
is denoted by $S^{\pm}$,
where the superscript denotes the $\mathbb{Z}_2$ parity.

\textit{Accessing quantum criticality through the state-operator correspondence.---}
We first assume the existence of a quantum critical point
$\delta u=\delta v=\delta h=0$
at which $\widehat{H}_{\mathrm{fzs}}(R)$
exhibits the conformal invariance of the
(2+1)D O(3) WF universality class
in the thermodynamic limit $R\propto\sqrt{N}\uparrow\infty$.
If we denote eigenstates of $\widehat{H}_{\mathrm{fzs}}(R)$
by the label $\mathrm{o}$ (see Table \ref{tab:bs_data_table}),
the state-operator
correspondence~\cite{Rychkov_2017}
implies that the eigenenergy $E_{\mathrm{o}}(R)$
measured relative to that of the ground state $E_{0}(R)$
scales like
\begin{align}
\delta E_{\mathrm{o}}(R)\equiv
E_{\mathrm{o}}(R)-E_{0}(R)\sim
\frac{c}{R}\,
\Delta_{\mathrm{o}}
\qquad
(\hbar=1),
\label{eq:deltaE_o at criticality}
\end{align}
in the limit $R\propto\sqrt{N}\uparrow\infty$.
Here, the radius $R$ of the fuzzy sphere is interpreted as
the radius of compactification in the radial quantization
of a CFT, $\Delta_{\mathrm{o}}$ the scaling dimension
of the CFT operator labeled by $\mathrm{o}$ and $c$
the system dependent speed of ``light'' in the CFT.

If we switch on a small perturbation
$\delta h=h-h_{\mathrm{c}}\neq0$
in $\widehat{H}_{\mathrm{fzs}}(R)$
holding $\delta u=\delta v=0$ fixed,
the energy shift to first-order in perturbation theory
is captured by adding the single relevant scalar operator
$\varepsilon$ (see Table \ref{tab:bs_data_table})
with the coupling $g_{\varepsilon}(h)$,
together with irrelevant operators
to the (2+1)D O(3) WF critical point.
Motivated by first-order conformal perturbation theory (CPT)%
~\cite{Icosahedron,Ising_CPT},
we make the scaling Ansatz
\begin{align}
\delta E_{\mathrm{o}}(R,h)=
\frac{c}{R}\,
\Delta_{\mathrm{o}}(R)
+
g_{\varepsilon}(h)\,
f_{\mathrm{o}\varepsilon\mathrm{o}}(R).
\label{eq:deltaE_o close to criticality}
\end{align}
Here 
$\Delta_{\mathrm{o}}(R)=\Delta_{\mathrm{o}}+\mathcal{O}(R^{-\omega})$
deviate from its value at quantum criticality due to the presence of
the leading irrelevant operator perturbing the CFT with the scaling exponent
$\omega>0$. The value of $h_{\mathrm{c}}$
at the quantum critical point is identified by
extrapolating the value of the root to $g_{\varepsilon}(h)=0$
to its limiting value when $R\uparrow\infty$.
The interpretation of 
$f_{\mathrm{o}\varepsilon\mathrm{o}}(R)=
f_{\mathrm{o}\varepsilon\mathrm{o}}+\mathcal{O}(R^{-\omega})$
in the limit $R\uparrow\infty$ is the following.
(i)
It is the (universal) OPE coefficient corresponding
to fusing operators
$\mathrm{o}$ and $\varepsilon$ into $\mathrm{o}$,
if $\mathrm{o}$ labels a scalar primary field.
(ii)
It is proportional to
$f_{\mathrm{o}_{\mathrm{p}}\varepsilon\mathrm{o}_{\mathrm{p}}}(R)$
with the propotionality constant a known function of 
$\Delta_{\mathrm{o}_{\mathrm{p}}}$ and $\Delta_{\varepsilon}$
if $\mathrm{o}$ labels the descendant of a primary field labeled by
$\mathrm{o}_{\mathrm{p}}$%
~\cite{Icosahedron,Ising_CPT}.
(iii) Finally,
it is a combination of OPE coefficients
~\cite{Ising_CPT}
if $\mathrm{o}$ labels a field with non-vanishing conformal spin.

Our strategy to establish the existence of
the (2+1)D O(3) WF critical point
from numerical diagonalization of $\widehat{H}_{\mathrm{fzs}}(R)$
proceeds in two steps. First, we choose two
eigenstates (CFT operators) of $\widehat{H}_{\mathrm{fzs}}(R)$ labeled by
$\mathrm{o}_{1}$
and
$\mathrm{o}_{2}$,
for which estimates of
$\Delta_{\mathrm{o}_{i}}$
and
$f_{\mathrm{o}_{i}\varepsilon\mathrm{o}_{i}}$
with $i=1,2$
at the (2+1)D O(3) WF critical point
are known, to solve for $c$ and $g_{\varepsilon}(h)$
from the pair of equations (\ref{eq:deltaE_o close to criticality})
they generate. 
Second, for any  eigenstate (CFT operator)
of $\widehat{H}_{\mathrm{fzs}}(R)$ labeled by
$\mathrm{o}$,
we extract $\Delta_{\mathrm{o}}(R)$ by using the extracted value of $c$ in
Eq.\ (\ref{eq:deltaE_o close to criticality})
with $g_{\varepsilon}(h)=0$
and then extract $f_{\mathrm{o}\varepsilon\mathrm{o}}$
from the dependence on $R$ of
Eq.\ (\ref{eq:deltaE_o close to criticality})
with $g_{\varepsilon}(h)\neq0$.

\textit{Results for $c$ and $h_{\mathrm{c}}$.---}
To determine $c$ and $g_{\varepsilon}(h)$,
we choose the pair of eigenstates labeled by
$\mathrm{o}_{1}\equiv\sigma$
and
$\mathrm{o}_{2}\equiv\partial_{\mu}\sigma$,
where $\sigma$ labels the lowest eigenenergy of $\widehat{H}_{\mathrm{fzs}}(R)$
with the quantum numbers $S=1^{-}$ and  $L=0$,
while $\partial_\mu\sigma$
labels the lowest eigenenergy of $\widehat{H}_{\mathrm{fzs}}(R)$
with the quantum numbers
$S=1^{-}$, $L=1$. In CFT jargon,
$\sigma$ labels the most relevant primary field and
$\partial_\mu\sigma$ is its first descendent.
We insert in 
Eq.\ (\ref{eq:deltaE_o close to criticality})
$\mathrm{o}_{1}\equiv\sigma$
and
$\mathrm{o}_{2}\equiv\partial_{\mu}\sigma$,
the conformal bootstrap (CB) estimates
$\Delta_\sigma=0.518936$
from Table \ref{tab:bs_data_table}
and
$\Delta_{\partial_\mu\sigma}=1+\Delta_\sigma=1.518936$
together with the CB estimates for
the OPE coefficient
$f_{\sigma\varepsilon\sigma}=0.525$
and
$f_{\partial_\mu\sigma\varepsilon\partial_\mu\sigma}=
f_{\sigma\varepsilon\sigma}\,\mathcal{A}_{\sigma,\varepsilon}$
with
$\mathcal{A}_{\sigma,\varepsilon}=
1+[\Delta_{\varepsilon}(\Delta_{\varepsilon}-3)/(6\Delta_{\sigma})]$
and $\Delta_{\varepsilon}$ given by its CB value quoted in
Table \ref{tab:bs_data_table}.
The factor $\mathcal{A}_{\sigma,\varepsilon}$ accounts for the fact
that the perturbation affects the first descendent differently from
the primary~\cite{Icosahedron,Ising_CPT}.  Solving these equations, we
obtain the speed of light $c\sim 0.0122$ and the coupling
$g_{\varepsilon}(h)$.  We then solve for the root of
$g_{\varepsilon}(h)=0$ for each value of $N$, to obtain the
finite-size critical point as shown in
Fig.~\ref{fig:critical-points}(a).  By extrapolating the root of
$g_{\varepsilon}(h)=0$ to $N\uparrow\infty$ as shown in
Fig.~\ref{fig:critical-points}(b), we obtain the critical point of the
system as $h_{\mathrm{c}}=14.992$.

\begin{figure*} % Use [t] or [b] for placement at top or bottom
\centering
\includegraphics[width=\textwidth]{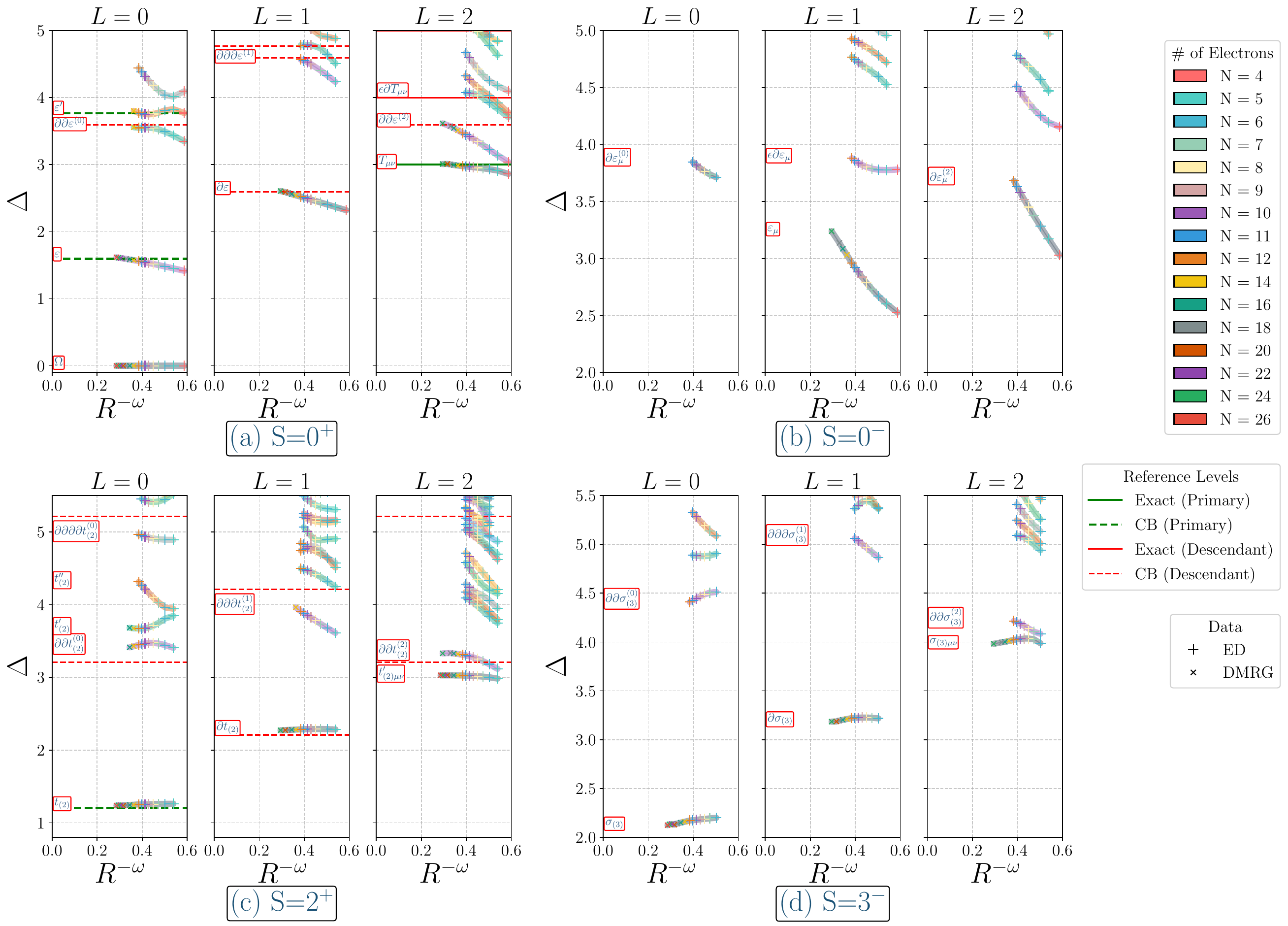}
\caption{
Dependence on $R^{-\omega}$ of the dimensionless number
$\Delta_{\mathrm{o}}(R)$
defined by the scaling Ansatz
(\ref{eq:deltaE_o close to criticality})
with $h$ chosen to be the root of $g_{\varepsilon}(h)=0$ for each value of $R$.
Here, the value $\omega\equiv\Delta_{\varepsilon'}-3\approx0.7668$
is deduced from the CB estimate for the scaling dimension
$\Delta_{\varepsilon'}$ 
of the leading irrelevant CFT perturbation
(see Table \ref{tab:bs_data_table}). The limiting value
$\Delta_{\mathrm{o}}$ of $\Delta_{\mathrm{o}}(R)$ as $R\uparrow\infty$
is interpreted as the scaling dimension of the CFT operator labeled by
$\mathrm{o}$.
The choice of the eigenstates labeled by $\mathrm{o}$ is organized by
$S^{\pm}$ sectors across panels; and by angular momenta $L$ within
each panel.
Dashed lines indicate conformal bootstrap predictions with the color code
green for primary operators and red for descendants.
Solid lines represent exact CFT values as applicable
to the stress-energy tensor and the conserved Noether current
(using the same color code). Operator
names, such as $\varepsilon$ and $\varepsilon_{\mu}$, are indicated in
red boxes. Only selected sectors are shown here; the complete spectra
are available in the Supplementary Material~\cite{SuppMat}. Panel (a)
displays the $S=0^{+}$ sector, panel (b) the $S=1^{+}$ sector, panel
(c) the $S=1^{-}$ sector, and panel (d) the $S=2^{+}$ sector.  Note
that, except for the $\sigma$ and $\partial_\mu \sigma$ levels used to
determine $c$ and $h_{\mathrm{c}}$, all the plotted values of 
$\Delta_{\mathrm{o}}(R)$ as $R\uparrow\infty$
are parameter‑free outputs.
         }
\label{fig:O3_spectra_grid}
\end{figure*}

\textit{Results for scaling dimensions.---}
The dependence on $R^{3-\Delta_{\varepsilon'}}$
of the dimensionless number
$\Delta_{\mathrm{o}}(R)$
defined by the scaling Ansatz
(\ref{eq:deltaE_o close to criticality})
with $h$ chosen to be the root of $g_{\varepsilon}(h)=0$ for each value of $R$
is shown in Fig.~\ref{fig:O3_spectra_grid} for a selection of
eigenstates of $\widehat{H}_{\mathrm{fzs}}(R)$ obtained from ED and DMRG.
Here, $\Delta_{\varepsilon'}$ is the scaling dimension of the leading
irrelevant perturbation, which is estimated to be $3.7668$ by CB.
Our estimates for the scaling dimensions
$\Delta_{\mathrm{o}}$, obtained from the largest available ED and DMRG system sizes rather than from an explicit extrapolation to $R \uparrow \infty$,
are reported in Table~\ref{tab:bs_data_table}
when $\mathrm{o}$ labels a primary operator
whose scaling dimension is known either from the conformal bootstrap or exactly,
and in Table~\ref{tab: primaries table}
(Appendix~\ref{sec: appendix Primary operators}) otherwise.
A systematic list for the values of the
scaling dimensions for all symmetry sectors
can be found in the supplementary material~\cite{SuppMat}.

Figure~\ref{fig:O3_spectra_grid}
reveals signatures of emergent conformal symmetry
that includes the expected descendant structure,
the stress energy tensor, and the conserved SO(3) Noether currents%
~\cite{Zhu_Han_Huffman_Hofmann_He_2023}.
The CB value of the scaling dimension
$\Delta_{\varepsilon}$
for the primary operator $\varepsilon$
corresponding to the quantum numbers
$S=0^{+}$ and $L=0$
is consistent with its ED lower bound
and DMRG upper bound from Table \ref{tab:bs_data_table},
respectively. The exact value of the scaling dimension
$\Delta_{T_{\mu\nu}}=3$
for the stress-energy tensor
$T_{\mu\nu}$
corresponding to the quantum numbers
$S=0^{+}$ and $L=2$
is consistent with the ED lower bound
and the DMRG upper bound from Table \ref{tab:bs_data_table}.
The exact value of the scaling dimension
$\Delta_{j_{\mu}}=2$
for the conserved SO(3) Noether current $j_{\mu}$
corresponding to the quantum numbers
$S=1^{+}$, $L=1$,
that reflects the internal SO(3) symmetry
is also consistent with the ED and DMRG estimates
from Table \ref{tab:bs_data_table}.
In addition, the CB predictions for the
scaling dimensions of the 
rank-2 $t_{(2)}$ and rank-4 $t_{(4)}$ O(3) primaries 
are consistent with
the ED and DMRG estimates from Table \ref{tab:bs_data_table}.
Finally, according to Fig.\ \ref{fig:O3_spectra_grid} and
Table \ref{tab: primaries table} (Appendix
\ref{sec: appendix Primary operators}),
there is no CB prediction for
the scaling dimension $\Delta_{\sigma_{(3)}}\approx2.124$
where $\sigma_{(3)}$ corresponds to the quantum numbers
$S=3^{-}$ [rank 3 O(3) pseudotensor]
and $L=0$ [SO(3) scalar].

In addition, a variety of irrelevant primaries appear, including dangerously
irrelevant~\cite{AMIT_dangerous_irrelevant_operators} ones. Among
these, the most significant are $\varepsilon'$,
a scalar under both O(3) and SO(3)
[see Fig.~\ref{fig:O3_spectra_grid}(a)],
and $\varepsilon_{\mu}$~\cite{Henriksson:2025hwi}
a pseudoscalar under O(3) and a pseudovector under SO(3)
[see Fig.~\ref{fig:O3_spectra_grid}(b)].
These operators govern leading corrections to scaling in lattice
simulations~\cite{Cubic_interactions_Fritz_Wessel_Vojta,Sandvik_2018}.
Notably, $\varepsilon_{\mu}$ provides insight
into a long-standing discrepancy in the critical exponents of columnar
versus staggered dimerized
antiferromagnets%
~\cite{Cubic_interactions_Fritz_Wessel_Vojta,Unconventional_Kao_Jiang_2012, Shinya_Synge_2013, Sandvik_2018},
once misattributed to different universality
classes~\cite{Unconventional_Wenzel_2008}.  

% Our finite-size
% extrapolation using a single power law with exponent $0.7668$ yields
% the scaling dimension
% $\Delta\approx3.931$,
% in reasonable agreement with the five-loop $\epsilon$-expansion result of
% 3.78796~\cite{Henriksson:2025hwi}.
% The identification of this operator as a
% three-body term with two derivatives and dimension 3.5 in the free
% O(3) theory further corroborates predictions in the
% literature~\cite{Cubic_interactions_Fritz_Wessel_Vojta}.

At last,
the spectrum of scaling exponent that we found numerically
is compared with those obtained from the large quantum-number or large-$S$ expansion
[see Refs.\ \cite{Cuomo2021,Cuomo2020, SuppMat}
and
Appendix \ref{sec:appendix large S expansion}],
i.e., a Laurent expansion in powers of $1/S$ of the scaling dimensions
of some low-lying primaries in the internal SO(3) symmetry sector
labeled by the quantum number $S\gg1$.
Our data agree with this large $S$ expansion, in particular for the
low-$L$ primaries.
They are consistent with the existence of both a 
gapless, linearly dispersing
phonon mode and of a dispersive gapped Goldstone mode.

\begin{figure}
\centering
\includegraphics[width=\linewidth]{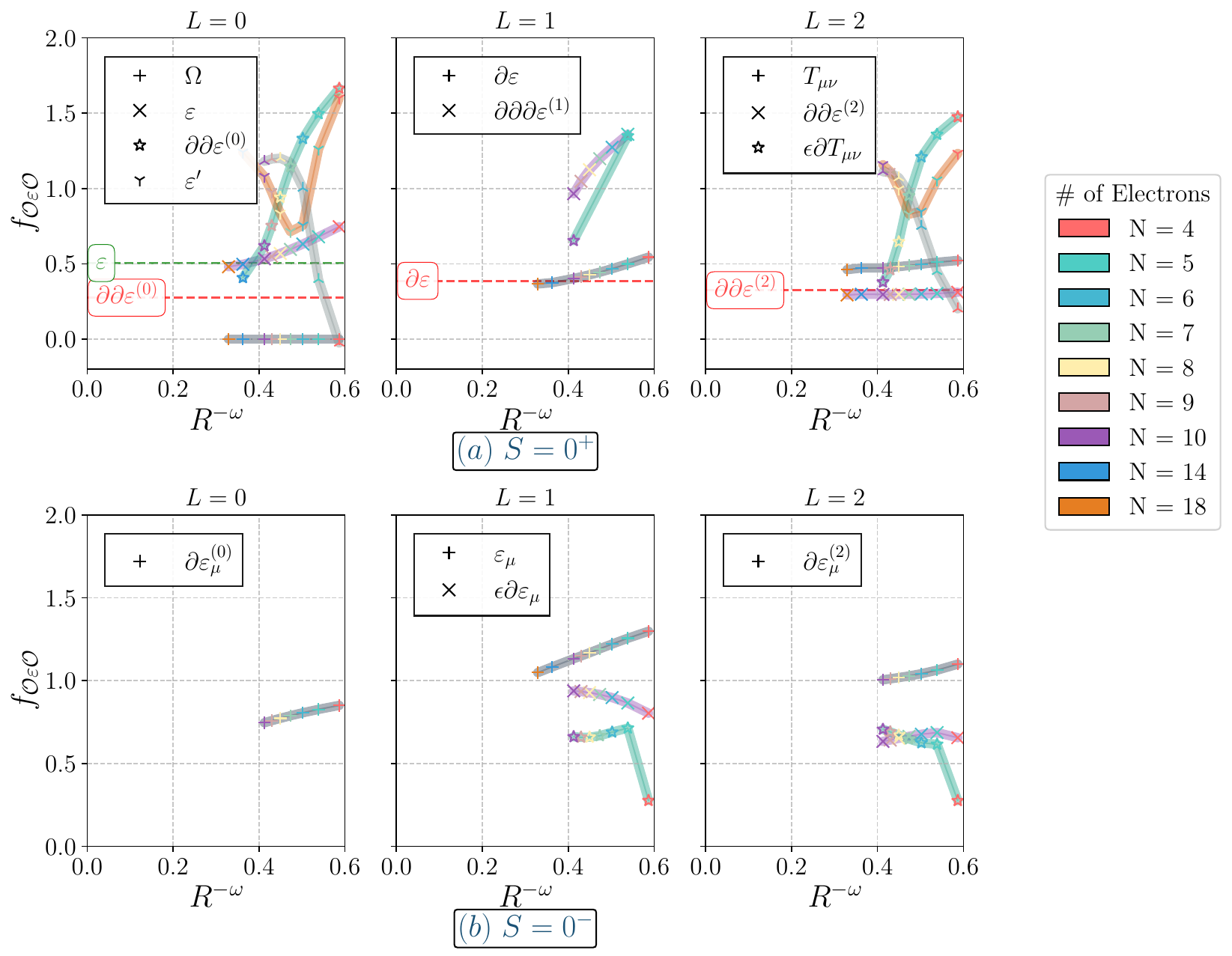}
\caption{
Dependence on $R^{-\omega}$ with $\omega\equiv\Delta_{\varepsilon'}-3\approx0.7668$
of the coefficient
$f_{\mathrm{o}\varepsilon\mathrm{o}}(R)=
f_{\mathrm{o}\varepsilon\mathrm{o}}
+
\mathcal{O}(R^{-\omega})$
defined by Eq.\ (\ref{eq:def fovarepsilono(R)})
with $\mathrm{o}$ chosen in the $S=0^{\pm}$
symmetry sectors. Solid lines denote CB estimates.
Dashed red lines include CPT-derived descendant factors applied to the
bootstrap data~\cite{Icosahedron}. The states (CFT operators)
selected by $\mathrm{o}$ include $\varepsilon$, $\varepsilon_{\mu}$,
$T_{\mu\nu}$.} 
\label{fig: main ope plot}
\end{figure}

\textit{Results for OPE coefficients.---}
We extract
$f_{\mathrm{o}\varepsilon\mathrm{o}}(R)$
in the scaling Ansatz (\ref{eq:deltaE_o close to criticality})
with the help of
\begin{equation}
f_{\mathrm{o}\varepsilon\mathrm{o}}(R)=
\left.
\frac{\partial\,\delta\,E_{\mathrm{o}}(R)}{\partial\,g_{\varepsilon}(R)}
\right|^{\,}_{g_{\varepsilon}(R)=0}.
\label{eq:def fovarepsilono(R)}
\end{equation}
In Fig.~\ref{fig: main ope plot},
we plot the dependence on $R^{-\omega}$ of
$f_{\mathrm{o}\varepsilon\mathrm{o}}(R)$
for some of these OPE coefficients and we compare their limiting values
as $R\uparrow\infty$ to the available CB data.
Other OPE coefficients are plotted
and tabulated in the Supplementary Material~\cite{SuppMat}.

\textit{Discussion and outlook.---}
We have mapped the square-lattice quantum rotor model, 
whose Hamiltonian exhibits a quantum critical point in the 
(2+1)-dimensional O(3) WF universality class, 
onto the fuzzy sphere Hamiltonian~(\ref{eq:def H fs}) 
by placing interacting fermions with four internal flavors 
on the sphere.
Our method remains computationally tractable
using ED and DMRG even for a maximum of $N=26$ interacting fermions
and provides evidences for an emergent
conformal symmetry consistent with the
(2+1)-dimensional O(3) WF universality class.
We identify 24 primary operators labeled by the index $\mathrm{o}$
in Table \ref{tab:bs_data_table} and Table
\ref{tab: primaries table} (Appendix
\ref{sec: appendix Primary operators})
and compute their scaling dimensions
$\Delta_{\mathrm{o}}$
as well as their
diagonal OPE coefficients $f_{\mathrm{o}\varepsilon\mathrm{o}}$.
The values that we find agrees with published ones and 
confirm predictions from the large-$S$
expansion~\cite{Monin17,Cuomo2020, Cuomo2021}
(see Appendix \ref{sec:appendix large S expansion}).

While analyzing the (2+1)-dimensional O(3) WF CFT,
we identified a weakly irrelevant
operator $\varepsilon_{\mu}$ relevant to the criticality of staggered
dimerized antiferromagnets. This operator transforms as a pseudoscalar
under O(3) and a pseudovector under Lorentz [SO(3)] symmetry, originating
from a cubic two-derivative term in the free theory, reminiscent of
the topological $\theta$-term.
This operator also admits a complementary description
as a local bulk interaction in the soft-spin $\phi^4$
framework~\cite{Cubic_interactions_Fritz_Wessel_Vojta},
and its topological character under lattice regularization
is not fully settled.
Regardless of interpretation, its weak irrelevance generates
anomalously large corrections to scaling
in staggered dimerized antiferromagnets,
supporting a single O(3) universality class
with strong subleading corrections
rather than a distinct universality class.
Boundary CFT provides a natural setting to
distinguish the two pictures, and our bulk identification
of $\varepsilon_{\mu}$ makes this test quantitatively accessible.

Our framework naturally generalizes to the study of $\mathrm{O}(N)$
WF quantum criticality, with explicit Hamiltonian constructions
provided in the supplementary material~\cite{SuppMat}. Furthermore, it
enables systematic exploration of models exhibiting richer internal
symmetries, such as
$\mathrm{O}(n_1)\times\mathrm{O}(n_2)$~\cite{Fisher_Bicritical_Tetracritical},
and by introducing anisotropic terms
that break $\mathrm{O}(n)$ symmetry down to a discrete hypercubic subgroup,
the stable cubic universality class for $n\geq3$.
The method is also compatible with boundary and defect
modifications of the fuzzy sphere, opening access to
boundary operator spectra and defect observables.
More broadly, the 24 primaries and OPE coefficients reported here
provide quantitative input for conformal bootstrap studies, control
corrections to scaling in lattice simulations, and constrain universal
features of experimental response functions near O(3) quantum criticality.

\begin{acknowledgments}
\textit{Acknowledgments.---}
The authors thank Frédéric Mila, Slava Rychkov and Gabriele Cuomo for useful
discussions.  LH\ acknowledges the Tremplin funding from CNRS
Physique.  AD\ acknowledges useful discussions with Jo\~ao
Penedones, Johan Henriksson and Giuseppe Mussardo. AD thanks IHES, Paris for hospitality, where these results were first presented by AD at the workshop "Fuzzy Sphere Meets Conformal Bootstrap", June 2-6, 2025.
\end{acknowledgments}

\textit{Note added.---}
While finishing this article, we became aware of an independent related work~\cite{guo2025onfreescalarwilsonfisherconformal}.

% References cited only in the Supplementary Material (included in the main
% bibliography as the last entries).
\nocite{Hellerman15,Alvarez-Gaume17,Banerjee18,Banerjee19,Alvarez-Gaume19,Watanabe21,Giombi21,Gaume21,Lange65,Nielsen76,Watanabe11,Watanabe12,Nicolis13,Watanabe13,ITensor,ITensor-r0.3,White1992,White2005}

\bibliographystyle{apsrev4-2} 
\bibliography{bibtex}

@article{Icosahedron,
  author = {Bing-Xin Lao and Slava Rychkov},
  title = {{3D Ising CFT and exact diagonalization on icosahedron: The power of conformal perturbation theory}},
  journal = {SciPost Phys.},
  publisher = {SciPost},
  volume = {15},
  pages = {243},
  year = {2023},
  doi = {10.21468/SciPostPhys.15.6.243},
  url = {https://scipost.org/10.21468/SciPostPhys.15.6.243},
  eprint = {2307.02540},
}

@article{Ising_CPT,
  author = {Andreas M. Läuchli and Loïc Herviou and Patrick H. Wilhelm and Slava Rychkov},
  title = {{Exact Diagonalization, Matrix Product States and Conformal Perturbation Theory Study of a 3D Ising Fuzzy Sphere Model}},
  journal = {SciPost Phys.},
  volume = {19},
  number = {3},
  pages = {076},
  year = {2025},
  doi = {10.21468/SciPostPhys.19.3.076},
  url = {https://arxiv.org/abs/2504.00842},
  eprint = {2504.00842},
}

@article{one_by_nu_hasenbuch,
  author = {Campostrini, Massimo and Hasenbusch, Martin and Pelissetto, Andrea and Rossi, Paolo and Vicari, Ettore},
  title = {Critical exponents and equation of state of the three-dimensional Heisenberg universality class},
  journal = {Phys. Rev. B},
  publisher = {American Physical Society},
  volume = {65},
  number = {14},
  pages = {144520},
  year = {2002},
  month = {Apr},
  doi = {10.1103/PhysRevB.65.144520},
  url = {https://link.aps.org/doi/10.1103/PhysRevB.65.144520},
  eprint = {cond-mat/0110336},
}

@article{HENRIKSSON20231,
  author = {Johan Henriksson},
  title = {The critical O(N) CFT: Methods and conformal data},
  journal = {Physics Reports},
  volume = {1002},
  pages = {1--72},
  year = {2023},
  doi = {10.1016/j.physrep.2022.12.002},
  url = {https://www.sciencedirect.com/science/article/pii/S0370157322004057},
  eprint = {2201.09520},
  issn = {0370-1573},
}

@article{Cuomo2021,
  author = {Cuomo, G.
and Esposito, A.
and Gendy, E.
and Khmelnitsky, A.
and Monin, A.
and Rattazzi, R.},
  title = {Gapped Goldstones at the cut-off scale: a non-relativistic EFT},
  journal = {Journal of High Energy Physics},
  volume = {2021},
  number = {2},
  pages = {68},
  year = {2021},
  month = {Feb},
  doi = {10.1007/JHEP02(2021)068},
  eprint = {2005.12924},
  issn = {1029-8479},
}

@article{O3Bootstrap_Shai,
  author = {Chester, Shai M. and Landry, Walter and Liu, Junyu and Poland, David and Simmons-Duffin, David and Su, Ning and Vichi, Alessandro},
  title = {Bootstrapping Heisenberg magnets and their cubic instability},
  journal = {Phys. Rev. D},
  publisher = {American Physical Society},
  volume = {104},
  number = {10},
  pages = {105013},
  year = {2021},
  month = {Nov},
  doi = {10.1103/PhysRevD.104.105013},
  url = {https://link.aps.org/doi/10.1103/PhysRevD.104.105013},
  eprint = {2011.14647},
}

@phdthesis{Cuomo2020,
  author = {Cuomo, Gabriel Francisco},
  title = {Large charge, semiclassics and superfluids: from broken symmetries to conformal field theories},
  school = {École Polytechnique Fédérale de Lausanne},
  address = {Lausanne, Switzerland},
  year = {2020},
  month = {September},
  url = {https://infoscience.epfl.ch/record/292506},
  note = {Thèse no. 8397, Faculté des sciences de base, Laboratoire de physique théorique des particules, Programme doctoral en physique},
}

@article{Zhu_Han_Huffman_Hofmann_He_2023,
  author = {Zhu, Wei and Han, Chao and Huffman, Emilie and Hofmann, Johannes S. and He, Yin-Chen},
  title = {Uncovering Conformal Symmetry in the 3D Ising Transition: State-Operator Correspondence from a Quantum Fuzzy Sphere Regularization},
  journal = {Physical Review X},
  volume = {13},
  number = {2},
  pages = {021009},
  year = {2023},
  month = {April},
  doi = {10.1103/PhysRevX.13.021009},
  url = {https://link.aps.org/doi/10.1103/PhysRevX.13.021009},
  eprint = {2210.13482},
  issn = {2160-3308},
  language = {en},
}

@article{Belavin_Polyakov_Zamolodchikov_1984,
  author = {Belavin, A.A. and Polyakov, A.M. and Zamolodchikov, A.B.},
  title = {Infinite conformal symmetry in two-dimensional quantum field theory},
  journal = {Nuclear Physics B},
  volume = {241},
  number = {2},
  pages = {333--380},
  year = {1984},
  month = {July},
  doi = {10.1016/0550-3213(84)90052-X},
  url = {https://linkinghub.elsevier.com/retrieve/pii/055032138490052X},
  issn = {05503213},
  language = {en},
}

@article{PismaZhETF,
  author = {Polyakov, Alexander M.},
  title = {{Conformal symmetry of critical fluctuations}},
  journal = {JETP Lett.},
  volume = {12},
  pages = {381--383},
  year = {1970},
}

@book{Cardy_1996,
  author = {Cardy, John},
  title = {Scaling and Renormalization in Statistical Physics},
  publisher = {Cambridge University Press},
  year = {1996},
  series = {Cambridge Lecture Notes in Physics},
}

@book{Rychkov_2017,
  author = {Rychkov, Slava},
  title = {{EPFL Lectures on Conformal Field Theory in D{\ensuremath{>}}= 3 Dimensions}},
  publisher = {Springer},
  year = {2016},
  month = {1},
  doi = {10.1007/978-3-319-43626-5},
  eprint = {1601.05000},
  isbn = {978-3-319-43625-8, 978-3-319-43626-5},
  reportnumber = {CERN-TH-2016-012},
  series = {SpringerBriefs in Physics},
}

@book{Di_Francesco_Mathieu_Senechal_1997,
  author = {Di Francesco, Philippe and Mathieu, Pierre and Sénéchal, David},
  title = {Conformal Field Theory},
  publisher = {Springer New York},
  address = {New York, NY},
  year = {1997},
  doi = {10.1007/978-1-4612-2256-9},
  url = {https://link.springer.com/10.1007/978-1-4612-2256-9},
  isbn = {978-1-4612-7475-9},
  series = {Graduate Texts in Contemporary Physics},
}

@book{goldenfeld2018lectures,
  author = {Goldenfeld, N.},
  title = {Lectures On Phase Transitions And The Renormalization Group},
  publisher = {CRC Press},
  year = {2018},
  url = {https://books.google.ch/books?id=pF0PEAAAQBAJ},
  isbn = {9780429962042},
}

@article{Witczak_Krempa_2012,
  author = {Witczak-Krempa, William and Ghaemi, Pouyan and Senthil, T. and Kim, Yong Baek},
  title = {Universal transport near a quantum critical Mott transition in two dimensions},
  journal = {Physical Review B},
  publisher = {American Physical Society (APS)},
  volume = {86},
  number = {24},
  year = {2012},
  month = {December},
  doi = {10.1103/physrevb.86.245102},
  eprint = {1206.3309},
  issn = {1550-235X},
}

@book{Sachdev_2011,
  author = {Sachdev, Subir},
  title = {Quantum Phase Transitions},
  publisher = {Cambridge University Press},
  year = {2011},
  edition = {2},
}

@article{Vichi_Slava_2011,
  author = {El-Showk, Sheer and Paulos, Miguel F. and Poland, David and Rychkov, Slava and Simmons-Duffin, David and Vichi, Alessandro},
  title = {Solving the 3D Ising model with the conformal bootstrap},
  journal = {Phys. Rev. D},
  publisher = {American Physical Society},
  volume = {86},
  number = {2},
  pages = {025022},
  year = {2012},
  month = {Jul},
  doi = {10.1103/PhysRevD.86.025022},
  url = {https://link.aps.org/doi/10.1103/PhysRevD.86.025022},
  eprint = {1203.6064},
}

@article{Hasenbusch_2022,
  author = {Hasenbusch, Martin},
  title = {Three-dimensional O ( N ) -invariant ϕ 4 models at criticality for N ≥ 4},
  journal = {Physical Review B},
  volume = {105},
  number = {5},
  pages = {054428},
  year = {2022},
  month = {February},
  doi = {10.1103/PhysRevB.105.054428},
  url = {https://link.aps.org/doi/10.1103/PhysRevB.105.054428},
  eprint = {2112.03783},
  issn = {2469-9950, 2469-9969},
  language = {en},
}

@article{Kos_Poland_Simmons-Duffin_Vichi_2016,
  author = {Kos, Filip and Poland, David and Simmons-Duffin, David and Vichi, Alessandro},
  title = {Precision islands in the Ising and O(N ) models},
  journal = {Journal of High Energy Physics},
  volume = {2016},
  number = {8},
  pages = {36},
  year = {2016},
  month = {August},
  doi = {10.1007/JHEP08(2016)036},
  url = {http://link.springer.com/10.1007/JHEP08(2016)036},
  eprint = {1603.04436},
  issn = {1029-8479},
}

@article{Derkachov_Manashov_1997_epsilon,
  author = {Derkachov, S. \'E. and Manashov, A. N.},
  title = {Stability Problem in the O(N) Nonlinear Sigma Model},
  journal = {Physical Review Letters},
  volume = {79},
  number = {8},
  pages = {1423--1427},
  year = {1997},
  month = {August},
  doi = {10.1103/PhysRevLett.79.1423},
  url = {https://link.aps.org/doi/10.1103/PhysRevLett.79.1423},
  eprint = {hep-th/9705020},
  issn = {0031-9007, 1079-7114},
}

@article{LANG1993573_1_N,
  author = {K. Lang and W. Rühl},
  title = {The critical O(N) σ-model at dimensions 2 < d < 4: a list of quasi-primary fields},
  journal = {Nuclear Physics B},
  volume = {402},
  number = {3},
  pages = {573--603},
  year = {1993},
  doi = {10.1016/0550-3213(93)90119-A},
  url = {https://www.sciencedirect.com/science/article/pii/055032139390119A},
  issn = {0550-3213},
}

@article{Conformal_Fuzzy_Wilson_Fisher_Content,
  author = {Han, Chao and Hu, Liangdong and Zhu, W.},
  title = {Conformal operator content of the Wilson-Fisher transition on fuzzy sphere bilayers},
  journal = {Phys. Rev. B},
  publisher = {American Physical Society},
  volume = {110},
  number = {11},
  pages = {115113},
  year = {2024},
  month = {Sep},
  doi = {10.1103/PhysRevB.110.115113},
  url = {https://link.aps.org/doi/10.1103/PhysRevB.110.115113},
  eprint = {2312.04047},
}

@misc{SuppMat,
  note = {See the Supplementary Material for additional figures, derivations, and computational details. The supplementary material includes the references at the end of this Reference list.},
}

@article{Cubic_interactions_Fritz_Wessel_Vojta,
  author = {Fritz, L. and Doretto, R. L. and Wessel, S. and Wenzel, S. and Burdin, S. and Vojta, M.},
  title = {Cubic interactions and quantum criticality in dimerized antiferromagnets},
  journal = {Phys. Rev. B},
  publisher = {American Physical Society},
  volume = {83},
  number = {17},
  pages = {174416},
  year = {2011},
  month = {May},
  doi = {10.1103/PhysRevB.83.174416},
  url = {https://link.aps.org/doi/10.1103/PhysRevB.83.174416},
}

@article{Sandvik_2018,
  author = {Ma, Nvsen and Weinberg, Phillip and Shao, Hui and Guo, Wenan and Yao, Dao-Xin and Sandvik, Anders W.},
  title = {Anomalous Quantum-Critical Scaling Corrections in Two-Dimensional Antiferromagnets},
  journal = {Phys. Rev. Lett.},
  publisher = {American Physical Society},
  volume = {121},
  number = {11},
  pages = {117202},
  year = {2018},
  month = {Sep},
  doi = {10.1103/PhysRevLett.121.117202},
  url = {https://link.aps.org/doi/10.1103/PhysRevLett.121.117202},
  eprint = {1804.01273},
}

@misc{Unconventional_Kao_Jiang_2012,
  author = {M. -T. Kao and D. -J. Tan and F. -J. Jiang},
  title = {Quantum phase transitions of 2-d dimerized spin-1/2 Heisenberg models with spatial anisotropy},
  year = {2012},
  url = {https://arxiv.org/abs/1202.1057},
  eprint = {1202.1057},
}

@article{Shinya_Synge_2013,
  author = {Yasuda, Shinya and Todo, Synge},
  title = {Monte Carlo simulation with aspect-ratio optimization: Anomalous anisotropic scaling in dimerized antiferromagnets},
  journal = {Phys. Rev. E},
  publisher = {American Physical Society},
  volume = {88},
  number = {6},
  pages = {061301},
  year = {2013},
  month = {Dec},
  doi = {10.1103/PhysRevE.88.061301},
  url = {https://link.aps.org/doi/10.1103/PhysRevE.88.061301},
  eprint = {1307.4529},
}

@article{Unconventional_Wenzel_2008,
  author = {Wenzel, Sandro and Bogacz, Leszek and Janke, Wolfhard},
  title = {Evidence for an Unconventional Universality Class from a Two-Dimensional Dimerized Quantum Heisenberg Model},
  journal = {Phys. Rev. Lett.},
  publisher = {American Physical Society},
  volume = {101},
  number = {12},
  pages = {127202},
  year = {2008},
  month = {Sep},
  doi = {10.1103/PhysRevLett.101.127202},
  url = {https://link.aps.org/doi/10.1103/PhysRevLett.101.127202},
}

@article{Fisher_Bicritical_Tetracritical,
  author = {Nelson, David R. and Kosterlitz, J. M. and Fisher, Michael E.},
  title = {Renormalization-Group Analysis of Bicritical and Tetracritical Points},
  journal = {Phys. Rev. Lett.},
  publisher = {American Physical Society},
  volume = {33},
  number = {14},
  pages = {813--817},
  year = {1974},
  month = {Sep},
  doi = {10.1103/PhysRevLett.33.813},
  url = {https://link.aps.org/doi/10.1103/PhysRevLett.33.813},
}

@article{Haldane_1983_FQHE,
  author = {Haldane, F. D. M.},
  title = {Fractional Quantization of the Hall Effect: A Hierarchy of Incompressible Quantum Fluid States},
  journal = {Physical Review Letters},
  volume = {51},
  number = {7},
  pages = {605--608},
  year = {1983},
  month = {August},
  doi = {10.1103/PhysRevLett.51.605},
  issn = {0031-9007},
  language = {en},
}

@article{ITensor,
  author = {Matthew Fishman and Steven R. White and E. Miles Stoudenmire},
  title = {{The ITensor Software Library for Tensor Network Calculations}},
  journal = {SciPost Phys. Codebases},
  pages = {4},
  year = {2022},
  doi = {10.21468/SciPostPhysCodeb.4},
  eprint = {2007.14822},
}

@article{ITensor-r0.3,
  author = {Matthew Fishman and Steven R. White and E. Miles Stoudenmire},
  title = {{Codebase release 0.3 for ITensor}},
  journal = {SciPost Phys. Codebases},
  pages = {4-r0.3},
  year = {2022},
  doi = {10.21468/SciPostPhysCodeb.4-r0.3},
}

@article{White1992,
  author = {White, Steven R.},
  title = {Density matrix formulation for quantum renormalization groups},
  journal = {Phys. Rev. Lett.},
  volume = {69},
  number = {19},
  pages = {2863--2866},
  year = {1992},
  doi = {10.1103/PhysRevLett.69.2863},
}

@article{White2005,
  author = {White, Steven R.},
  title = {Density matrix renormalization group algorithms with a single center site},
  journal = {Phys. Rev. B},
  volume = {72},
  number = {18},
  pages = {180403},
  year = {2005},
  doi = {10.1103/PhysRevB.72.180403},
  eprint = {cond-mat/0508709},
}

@article{AMIT_dangerous_irrelevant_operators,
  author = {Daniel J Amit and Luca Peliti},
  title = {On dangerous irrelevant operators},
  journal = {Annals of Physics},
  volume = {140},
  number = {2},
  pages = {207--231},
  year = {1982},
  doi = {10.1016/0003-4916(82)90159-2},
  url = {https://www.sciencedirect.com/science/article/pii/0003491682901592},
  issn = {0003-4916},
}

@article{Virasoro_1970,
  author = {Virasoro, M. A.},
  title = {Subsidiary Conditions and Ghosts in Dual-Resonance Models},
  journal = {Phys. Rev. D},
  publisher = {American Physical Society},
  volume = {1},
  number = {10},
  pages = {2933--2936},
  year = {1970},
  month = {May},
  doi = {10.1103/PhysRevD.1.2933},
  url = {https://link.aps.org/doi/10.1103/PhysRevD.1.2933},
}

@book{Takahashi_1999,
  author = {Takahashi, Minoru},
  title = {Thermodynamics of One-Dimensional Solvable Models},
  publisher = {Cambridge University Press},
  year = {1999},
}

@article{JMadore1992,
  author = {J. Madore},
  title = {The fuzzy sphere},
  journal = {Classical and Quantum Gravity},
  volume = {9},
  number = {1},
  pages = {69},
  year = {1992},
  month = {jan},
  doi = {10.1088/0264-9381/9/1/008},
  eprint = {hep-th/0101189},
}

@article{Fuzzy_OPE,
  author = {Hu, Liangdong and He, Yin-Chen and Zhu, W.},
  title = {Operator Product Expansion Coefficients of the 3D Ising Criticality via Quantum Fuzzy Spheres},
  journal = {Phys. Rev. Lett.},
  publisher = {American Physical Society},
  volume = {131},
  number = {3},
  pages = {031601},
  year = {2023},
  month = {Jul},
  doi = {10.1103/PhysRevLett.131.031601},
  url = {https://link.aps.org/doi/10.1103/PhysRevLett.131.031601},
  eprint = {2303.08844},
}

@article{Fuzzy_4pt_correlators,
  author = {Han, Chao and Hu, Liangdong and Zhu, W. and He, Yin-Chen},
  title = {Conformal four-point correlators of the three-dimensional Ising transition via the quantum fuzzy sphere},
  journal = {Phys. Rev. B},
  publisher = {American Physical Society},
  volume = {108},
  number = {23},
  pages = {235123},
  year = {2023},
  month = {Dec},
  doi = {10.1103/PhysRevB.108.235123},
  url = {https://link.aps.org/doi/10.1103/PhysRevB.108.235123},
  eprint = {2306.04681},
}

@article{Fuzzy_Defects,
  author = {Liangdong Hu and Yin-Chen He and W. Zhu},
  title = {Solving conformal defects in 3D conformal field theory using fuzzy sphere regularization},
  journal = {Nature Communications},
  volume = {15},
  number = {1},
  pages = {3659},
  year = {2024},
  doi = {10.1038/s41467-024-47978-y},
  eprint = {2308.01903},
  issn = {2041-1723},
}

@article{Fuzzy_g_function,
  author = {Zheng Zhou and Davide Gaiotto and Yin-Chen He and Yijian Zou},
  title = {{The $g$-function and defect changing operators from wavefunction overlap on a fuzzy sphere}},
  journal = {SciPost Phys.},
  publisher = {SciPost},
  volume = {17},
  pages = {021},
  year = {2024},
  doi = {10.21468/SciPostPhys.17.1.021},
  url = {https://scipost.org/10.21468/SciPostPhys.17.1.021},
  eprint = {2401.00039},
}

@article{Fuzzy_Entropic_F_function,
  author = {Hu, Liangdong and Zhu, W. and He, Yin-Chen},
  title = {Entropic $F$ function of three-dimensional Ising conformal field theory via fuzzy sphere regularization},
  journal = {Phys. Rev. B},
  publisher = {American Physical Society},
  volume = {111},
  number = {15},
  pages = {155151},
  year = {2025},
  month = {Apr},
  doi = {10.1103/PhysRevB.111.155151},
  url = {https://link.aps.org/doi/10.1103/PhysRevB.111.155151},
  eprint = {2401.17362},
}

@article{Fuzzy_Impurities,
  author = {Gabriel Cuomo and Yin-Chen He and Zohar Komargodski},
  title = {Impurities with a cusp: general theory and 3d Ising},
  journal = {Journal of High Energy Physics},
  volume = {2024},
  number = {11},
  pages = {61},
  year = {2024},
  doi = {10.1007/JHEP11(2024)061},
  eprint = {2406.10186},
  issn = {1029-8479},
}

@article{Fuzzy_Surface_CFTs,
  author = {Zheng Zhou and Yijian Zou},
  title = {{Studying the 3d Ising surface CFTs on the fuzzy sphere}},
  journal = {SciPost Phys.},
  publisher = {SciPost},
  volume = {18},
  pages = {031},
  year = {2025},
  doi = {10.21468/SciPostPhys.18.1.031},
  url = {https://scipost.org/10.21468/SciPostPhys.18.1.031},
  eprint = {2407.15914},
}

@article{Fuzzy_Generators,
  author = {Giulia Fardelli and Andrew Liam Fitzpatrick and Emanuel Katz},
  title = {{Constructing the infrared conformal generators on the fuzzy sphere}},
  journal = {SciPost Phys.},
  publisher = {SciPost},
  volume = {18},
  pages = {086},
  year = {2025},
  doi = {10.21468/SciPostPhys.18.3.086},
  url = {https://scipost.org/10.21468/SciPostPhys.18.3.086},
  eprint = {2409.02998},
}

@article{Fuzzy_SO5,
  author = {Zhou, Zheng and Hu, Liangdong and Zhu, W. and He, Yin-Chen},
  title = {SO(5) Deconfined Phase Transition under the Fuzzy-Sphere Microscope: Approximate Conformal Symmetry, Pseudo-Criticality, and Operator Spectrum},
  journal = {Phys. Rev. X},
  publisher = {American Physical Society},
  volume = {14},
  number = {2},
  pages = {021044},
  year = {2024},
  month = {Jun},
  doi = {10.1103/PhysRevX.14.021044},
  url = {https://link.aps.org/doi/10.1103/PhysRevX.14.021044},
  eprint = {2306.16435},
}

@misc{FuzzyRealScalarHe,
  author = {He, Yin-Chen},
  title = {Free real scalar CFT on fuzzy sphere: spectrum, algebra and wavefunction ansatz},
  year = {2025},
  month = {June},
  doi = {10.48550/arXiv.2506.14904},
  url = {http://arxiv.org/abs/2506.14904},
  eprint = {2506.14904},
}

@article{FuzzyRealScalarTaylor,
  author = {Taylor, Joseph and Voinea, Cristian and Papić, Zlatko and Fan, Ruihua},
  title = {Conformal scalar field theory from Ising tricriticality on the fuzzy sphere},
  journal = {Phys. Rev. Lett.},
  volume = {136},
  number = {5},
  pages = {056503},
  year = {2026},
  month = {June},
  doi = {10.48550/arXiv.2506.22539},
  url = {http://arxiv.org/abs/2506.22539},
  eprint = {2506.22539},
}

@article{Hellerman15,
  author = {{Hellerman}, Simeon and {Orlando}, Domenico and {Reffert}, Susanne and {Watanabe}, Masataka},
  title = {{On the CFT operator spectrum at large global charge}},
  journal = {Journal of High Energy Physics},
  volume = {2015},
  pages = {71},
  year = {2015},
  month = {December},
  doi = {10.1007/JHEP12(2015)071},
  eprint = {1505.01537},
}

@article{Alvarez-Gaume17,
  author = {{Alvarez-Gaume}, Luis and {Loukas}, Orestis and {Orlando}, Domenico and {Reffert}, Susanne},
  title = {{Compensating strong coupling with large charge}},
  journal = {Journal of High Energy Physics},
  volume = {2017},
  number = {4},
  pages = {59},
  year = {2017},
  month = {April},
  doi = {10.1007/JHEP04(2017)059},
  eprint = {1610.04495},
}

@article{Monin17,
  author = {{Monin}, A. and {Pirtskhalava}, D. and {Rattazzi}, R. and {Seibold}, F.~K.},
  title = {{Semiclassics, Goldstone bosons and CFT data}},
  journal = {Journal of High Energy Physics},
  volume = {2017},
  number = {6},
  pages = {11},
  year = {2017},
  month = {June},
  doi = {10.1007/JHEP06(2017)011},
  eprint = {1611.02912},
}

@article{Banerjee18,
  author = {{Banerjee}, Debasish and {Chandrasekharan}, Shailesh and {Orlando}, Domenico},
  title = {{Conformal Dimensions via Large Charge Expansion}},
  journal = {\prl},
  volume = {120},
  number = {6},
  pages = {061603},
  year = {2018},
  month = {February},
  doi = {10.1103/PhysRevLett.120.061603},
  eprint = {1707.00711},
}

@article{Banerjee19,
  author = {{Banerjee}, Debasish and {Chandrasekharan}, Shailesh and {Orlando}, Domenico and {Reffert}, Susanne},
  title = {{Conformal Dimensions in the Large Charge Sectors at the O (4 ) Wilson-Fisher Fixed Point}},
  journal = {\prl},
  volume = {123},
  number = {5},
  pages = {051603},
  year = {2019},
  month = {August},
  doi = {10.1103/PhysRevLett.123.051603},
  eprint = {1902.09542},
}

@article{Alvarez-Gaume19,
  author = {Alvarez-Gaume, Luis and Orlando, Domenico and Reffert, Susanne},
  title = {{Large charge at large N}},
  journal = {JHEP},
  volume = {12},
  number = {12},
  pages = {142},
  year = {2019},
  doi = {10.1007/JHEP12(2019)142},
  eprint = {1909.02571},
}

@article{Watanabe21,
  author = {Watanabe, Masataka},
  title = {{Accessing large global charge via the $\epsilon$-expansion}},
  journal = {JHEP},
  volume = {04},
  number = {4},
  pages = {264},
  year = {2021},
  doi = {10.1007/JHEP04(2021)264},
  eprint = {1909.01337},
}

@article{Giombi21,
  author = {Giombi, Simone and Hyman, Jonah},
  title = {{On the large charge sector in the critical O(N) model at large N}},
  journal = {JHEP},
  volume = {09},
  number = {9},
  pages = {184},
  year = {2021},
  doi = {10.1007/JHEP09(2021)184},
  eprint = {2011.11622},
}

@article{Gaume21,
  author = {Gaum{\'e}, Luis {\'A}lvarez and Orlando, Domenico and Reffert, Susanne},
  title = {{Selected topics in the large quantum number expansion}},
  journal = {Phys. Rept.},
  volume = {933},
  pages = {1--66},
  year = {2021},
  doi = {10.1016/j.physrep.2021.08.001},
  eprint = {2008.03308},
}

@article{Lange65,
  author = {Lange, R. V.},
  title = {Goldstone Theorem in Nonrelativistic Theories},
  journal = {Phys. Rev. Lett.},
  publisher = {American Physical Society},
  volume = {14},
  number = {1},
  pages = {3--6},
  year = {1965},
  month = {Jan},
  doi = {10.1103/PhysRevLett.14.3},
  url = {https://link.aps.org/doi/10.1103/PhysRevLett.14.3},
}

@article{Nielsen76,
  author = {H.B. Nielsen and S. Chadha},
  title = {On how to count Goldstone bosons},
  journal = {Nuclear Physics B},
  volume = {105},
  number = {3},
  pages = {445--453},
  year = {1976},
  doi = {10.1016/0550-3213(76)90025-0},
  url = {https://www.sciencedirect.com/science/article/pii/0550321376900250},
  issn = {0550-3213},
}

@article{Watanabe11,
  author = {Watanabe, Haruki and Brauner, Tom\'a\ifmmode \check{s}\else \v{s}\fi{}},
  title = {Number of Nambu-Goldstone bosons and its relation to charge densities},
  journal = {Phys. Rev. D},
  publisher = {American Physical Society},
  volume = {84},
  number = {12},
  pages = {125013},
  year = {2011},
  month = {Dec},
  doi = {10.1103/PhysRevD.84.125013},
  url = {https://link.aps.org/doi/10.1103/PhysRevD.84.125013},
  eprint = {1109.6327},
}

@article{Watanabe12,
  author = {Watanabe, Haruki and Brauner, Tom\'a\ifmmode \check{s}\else \v{s}\fi{}},
  title = {Spontaneous breaking of continuous translational invariance},
  journal = {Phys. Rev. D},
  publisher = {American Physical Society},
  volume = {85},
  number = {8},
  pages = {085010},
  year = {2012},
  month = {Apr},
  doi = {10.1103/PhysRevD.85.085010},
  url = {https://link.aps.org/doi/10.1103/PhysRevD.85.085010},
  eprint = {1112.3890},
}

@article{Nicolis13,
  author = {Nicolis, Alberto and Piazza, Federico},
  title = {Implications of Relativity on Nonrelativistic Goldstone Theorems: Gapped Excitations at Finite Charge Density},
  journal = {Phys. Rev. Lett.},
  publisher = {American Physical Society},
  volume = {110},
  number = {1},
  pages = {011602},
  year = {2013},
  month = {Jan},
  doi = {10.1103/PhysRevLett.110.011602},
  url = {https://link.aps.org/doi/10.1103/PhysRevLett.110.011602},
  eprint = {1204.1570},
}

@article{Watanabe13,
  author = {Watanabe, Haruki and Brauner, Tom\'a\ifmmode \check{s}\else \v{s}\fi{} and Murayama, Hitoshi},
  title = {Massive Nambu-Goldstone Bosons},
  journal = {Phys. Rev. Lett.},
  publisher = {American Physical Society},
  volume = {111},
  number = {2},
  pages = {021601},
  year = {2013},
  month = {Jul},
  doi = {10.1103/PhysRevLett.111.021601},
  url = {https://link.aps.org/doi/10.1103/PhysRevLett.111.021601},
  eprint = {1303.1527},
}

@article{Henriksson:2025hwi,
  author = {Johan Henriksson and Franz Herzog and Stefanos R. Kousvos and Jasper Roosmale Nepveu},
  title = {Multi-loop spectra in general scalar EFTs and CFTs},
  journal = {Phys.Lett.B},
  volume = {874},
  pages = {140235},
  year = {2026},
  doi = {10.1016/j.physletb.2026.140235},
  eprint = {2507.12518},
}

@misc{guo2025onfreescalarwilsonfisherconformal,
  author = {Wenhan Guo and Zheng Zhou and Tzu-Chieh Wei and Yin-Chen He},
  title = {The $O(N)$ Free-Scalar and Wilson-Fisher Conformal Field Theories on the Fuzzy Sphere},
  year = {2025},
  url = {https://arxiv.org/abs/2512.02234},
  eprint = {2512.02234},
}

\newpage

\section*{End Matter}

\subsection{Primary operators}
\label{sec: appendix Primary operators}

Table~\ref{tab: primaries table} presents the primary operators whose
scaling dimensions we have computed, for which conformal bootstrap
results are not currently available.

\begin{table}[htbp]
\centering
\caption{
Selected primary operators with the same conventions as
in Table \ref{tab:bs_data_table}.
        }
\begin{tabular}{|ccc|c||c|c|}
\hline
$S$ & $L$ & $I$ & $\mathrm{o}$ & $\Delta$(ED) & $\Delta$(DMRG) \\
\hline
$0^{-}$ & 1 & 1 & $\varepsilon_{\mu}$ & 2.96067 (12) & 3.2396 (18) \\
\hline
$1^{+}$ & 1 & 3 & $\phi_{\mu}$ & 3.70828 (12) & 3.75703 (18) \\
\hline
$1^{-}$ & 1 & 2 & $\sigma_{\mu}$ & 2.87055 (12) & 3.04373 (26) \\
\hline
$1^{-}$ & 2 & 2 & $\sigma_{\mu\nu}$ & 3.3557 (12) & 3.5937 (24) \\
\hline
$2^{+}$ & 0 & 3 & $t_{(2)}'$ & 3.67128 (12) & 3.68121 (16) \\
\hline
$2^{+}$ & 0 & 4 & $t_{(2)}''$ & 4.31261 (12) & -\\
\hline
$2^{+}$ & 2 & 1 & $t'_{(2)\mu\nu}$ & 3.02538 (12) & 3.02825 (26) \\
\hline
$2^{-}$ & 1 & 1 & $t_{(2)\mu}$ & 2.68970 (12) & 2.78187 (26) \\
\hline
$2^{-}$ & 2 & 1 & $t'_{(2)\mu\nu}$ & 3.29489 (12) & 3.5341 (24) \\
\hline
$3^{+}$ & 1 & 1 & $\chi_{(3)\mu}$ & 3.68769 (12) & 3.75082 (26) \\
\hline
$3^{+}$ & 2 & 1 & $\chi_{(3)\mu\nu}$ & 4.22326 (12) & -\\
\hline
$3^{-}$ & 0 & 1 & $\sigma_{(3)}$ & 2.17310 (12) & 2.12486 (26) \\
\hline
$3^{-}$ & 0 & 3 & & 4.88895 (11) & -\\
\hline
$3^{-}$ & 2 & 1 & $\sigma_{(3)\mu\nu}$ & 4.02221 (12) & 3.9825 (24) \\
\hline
$4^{+}$ & 0 & 3 & & 6.25369 (11) & -\\
\hline
$4^{+}$ & 2 & 1 & $t_{(4)\mu\nu}$ & 5.17757 (11) & -\\
\hline
$4^{-}$ & 1 & 1 & & 4.85139 (11) & 4.87182 (26) \\
\hline
$4^{-}$ & 2 & 1 & & 5.29724 (11) & -\\
\hline
\end{tabular}

\label{tab: primaries table}
\end{table}

\begin{figure}[t]
\centering
\includegraphics[width=\linewidth]{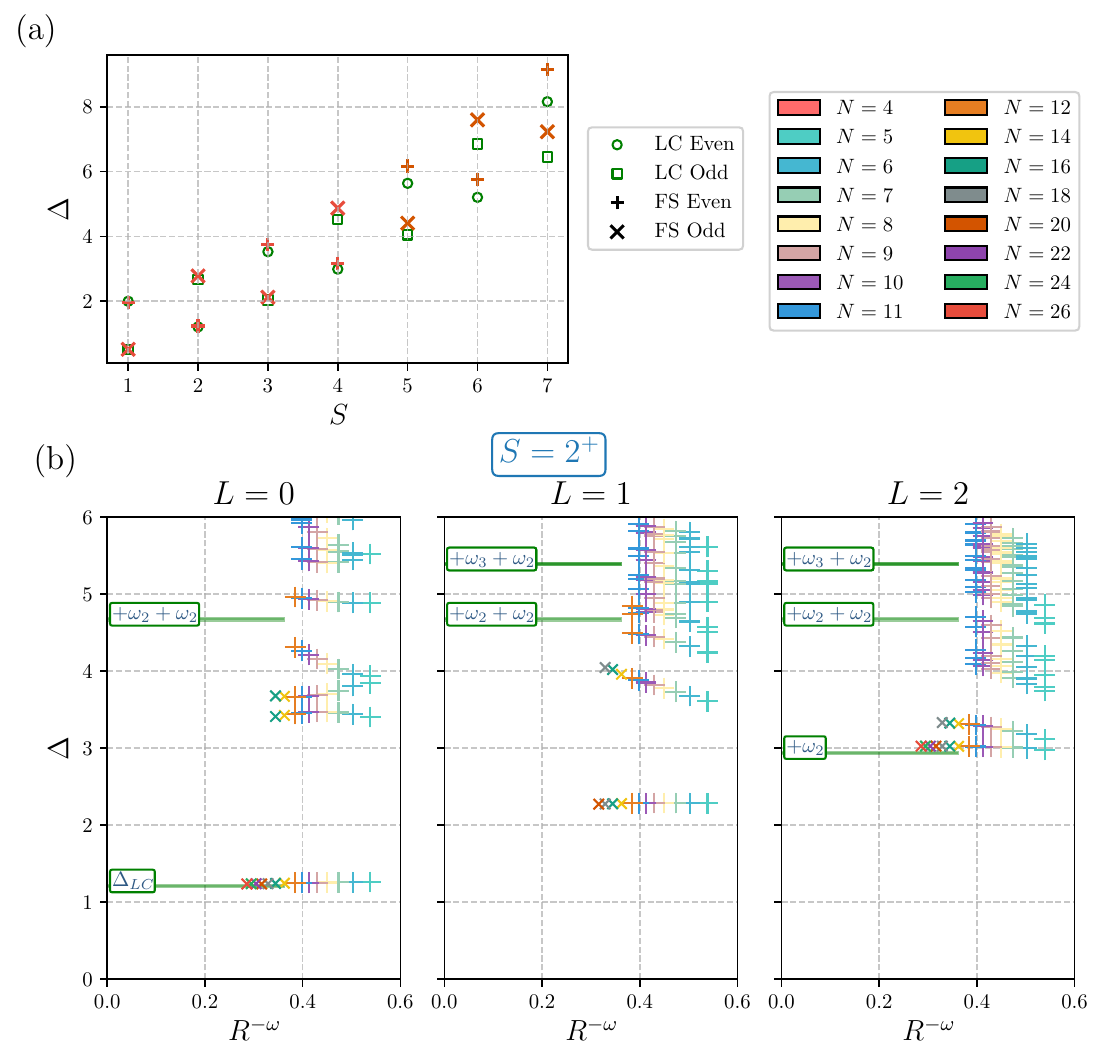}
\caption{
(a)
For each quantum number $S=1,2,\cdots$,
two scaling dimensions are reported.
The lowest one corresponds to the CFT state
in the symmetry sector with the spin quantum number
$S^{\kappa(S)}=1^{-},2^{+},\cdots$ with $\kappa(S)=(-1)^{S}$
for the internal symmetry group
$\mathrm{O}(3)=\mathbb{Z}_{2}\times\mathrm{SO}(3)$
and the angular momenta $L=0$.
The larger one corresponds to the CFT state
with the quantum numbers
$S^{\kappa(S+1)}=1^{+},2^{-},\cdots$
and
$L=1$.
The scaling dimensions obtained from ED and DMRG
of $\widehat{H}_{\mathrm{fzs}}(R)$
and reported as $+(\times)$ for $\kappa(S)=+(-)$.
For fuzzy sphere results, number of electrons is 26 (DMRG) for
$S=1,2,3,4$ and 12 (ED) for $S=5,6,7$.
The scaling dimensions obtained
from fitting the large-$S$ expansions
(\ref{eq:appendix large S expansion lowest scalar primaries})
and
(\ref{eq:appendix large charge expansion odd operator})
with bootstrap data
are reported as $\bigcirc(\square)$ for $\kappa(S)=+(-)$.
(b)  
Scaling dimensions as a function of
$R^{-\omega}$ with $\omega\approx0.7668$
for phonon primaries and their descendants
obtained from ED of $\widehat{H}_{\mathrm{fzs}}(R)$
in the symmetry sector $S=2^{+}$. 
The green lines report the scaling dimensions obtained
from fitting the large-$S$ expansion
(\ref{eq:appendix large charge expansion phonon modes})
with bootstrap data.
        }
\label{fig: large charge comparison}
\end{figure}

\subsection{Large-$S$ expansion}
\label{sec:appendix large S expansion}

The large-$S$ expansion provides a controlled semiclassical framework
for computing the scaling dimensions of local operators in a CFT
with internal O(3) symmetry when these local operators
transform according to the $2S+1$-dimensional
irreducible representation of O(3).
In the sector of the CFT with $S\gg1$,
some scaling dimensions can be associated with the dispersions of two
weakly interacting fields associated to the explicit symmetry breaking
of the internal symmetry SO(3) down to SO(2). One is massless and called
a phonon mode. The other is gapful and called a gapped Goldstone mode.

Three key predictions from the large-$S$ expansion
are the following when spacetime is three dimensional.

\textbf{1. Lowest scalar primaries.}
The scaling dimension of the lowest scalar $(L=0$) operator
in each $S\gg1$ sector obeys the semiclassical expansion
\begin{small}
\begin{equation}
\begin{split}
\Delta_0(S^{\kappa(S)},L=0)=&\,
\alpha\,
S^{3/2}
+
\beta\,
S^{1/2}
-
0.0937256
+
\gamma\,
S^{-1/2} \\
&\,
+
\mathcal{O}(1/S),
\end{split}
\label{eq:appendix large S expansion lowest scalar primaries}
\end{equation}
\end{small}

\noindent
where $\kappa(S) = (-1)^S$ determines the parity.
The Wilson coefficients
$\alpha = 0.31076$,
$\beta = 0.29818$,
and
$\gamma = 0.00370$
are obtained by fitting the lowest level in the sectors
$1^-$, $2^+$, $4^+$ to bootstrap data for $\sigma$, $t_{(2)}$, and $t_{(4)}$.

\textbf{2. Gapped Goldstone modes.} 
The scaling dimensions of the state
with spin $S\gg1$, parity $\kappa(S+1)=-\kappa(S)$, and
non-vanishing angular momenta $L\geq1$
obeys the semiclassical expansion
\begin{subequations}
\label{eq:appendix large charge expansion odd operator}
\begin{equation}
\begin{split}
\Delta(S^{\kappa(S+1)},L)=&\,
\Delta_0(S^{\kappa(S)},L=0)
+
\mu(S)
+
\chi\,
\frac{L(L+1)}{2\mu(S)}
\\
&\,
+
\mathcal{O}\left(\frac{L^4}{\mu^3}\right),
\end{split}
\label{eq:appendix large charge expansion odd operator a}
\end{equation}
where
\begin{equation}
\mu(S):=
\frac{\partial\Delta_0(S,L=0)}{\partial S}
\label{eq:appendix large charge expansion odd operator b}
\end{equation}
\end{subequations}  
is the chemical potential that breaks explicitly the internal
continuous symmetry from SO(3) to SO(2).
The term $L(L+1)/(2\mu)$ encodes the dispersion
of the gapped Goldstone boson. The value of the Wilson coefficient
$\chi$ is obtained by fitting the lowest scaling dimension in the sector
$1^+$ to the value 2 of the scaling dimension of the Noether current $j_{\mu}$.

\textbf{3. Gapless phonon modes.}
The scaling dimension of the state with the quantum numbers
$S\gg1$ for the spin, parity $\kappa(S)=(-1)^{S}$,
and the angular momenta $L=0,1,\cdots$
obeys the semiclassical expansion
\begin{subequations}
\label{eq:appendix large charge expansion phonon modes}
\begin{equation}
\Delta_0(S^{\kappa(S)},L)=
\Delta_0(S^{\kappa(S)},L=0)
+
\sum^{\,}_{L} n^{\,}_{L} \omega^{\,}_{L},
\label{eq:appendix large charge expansion phonon modes a}
\end{equation}
where 
\begin{equation}
\omega^{\,}_{L} = \sqrt{L(L+1)/2},
\label{eq:appendix large charge expansion phonon modes b}
\end{equation}
\end{subequations}
is the energy of the phonon with angular momenta $L$
and the integer $n^{\,}_{L}$ is its occupancy. The value 
$L=1$ correspond to a phonon descendant, while the values
$L\geq 2$ correspond to phonon primaries.

The agreement between these theoretical predictions and fuzzy sphere
numerical results (shown in Fig.~\ref{fig: large charge comparison})
provides strong evidence for the presence of both gapped Goldstone
modes and gapless phonon modes in the
(2+1)-dimensional O(3) WF CFT,
validating the large-$S$ expansion framework.

\end{document}

\title{Supplemental Material for:\\[1mm] \textit{Conformal Data for the O(3) Wilson-Fisher CFT from Fuzzy Sphere Realization of Quantum Rotor Model}}

% \title{Conformal Data for the O(3) Wilson-Fisher CFT from Fuzzy Sphere Realization of Quantum Rotor Model}

\author{Arjun Dey}
\affiliation{
Laboratory for Theoretical and Computational Physics,
PSI Center for Scientific Computing, Theory and Data,
5232 Villigen PSI, Switzerland}
\affiliation{
Institute of Physics,
\'{E}cole Polytechnique F\'{e}d\'{e}rale de Lausanne (EPFL),
1015 Lausanne, Switzerland}

\author{Loic Herviou}
\affiliation{Univ. Grenoble Alpes, CNRS, LPMMC, 38000 Grenoble, France}

\author{Christopher Mudry}
\affiliation{
Laboratory for Theoretical and Computational Physics,
PSI Center for Scientific Computing, Theory and Data,
5232 Villigen PSI, Switzerland}
\affiliation{
Institute of Physics,
\'{E}cole Polytechnique F\'{e}d\'{e}rale de Lausanne (EPFL),
1015 Lausanne, Switzerland}

\author{Andreas Martin L\"auchli} %{Andreas Martin L\"auchli Herzig}
\affiliation{
Laboratory for Theoretical and Computational Physics,
PSI Center for Scientific Computing, Theory and Data,
5232 Villigen PSI, Switzerland}
\affiliation{
Institute of Physics,
\'{E}cole Polytechnique F\'{e}d\'{e}rale de Lausanne (EPFL),
1015 Lausanne, Switzerland}

\maketitle
\tableofcontents
\clearpage

\section{Hamiltonian}

\noindent
We begin by recalling the Hamiltonian as defined in the main text:
\begin{equation}
\widehat{H}_{\mathrm{fzs}} :=
\widehat{P}_{\mathrm{LLL}}
\left(
u\,
\widehat{H}_{\mathrm{Hub}}
-
v\,
\widehat{H}_{\mathrm{Heis}}
+
h\,
\widehat{H}_{\mathrm{trans}}
\right)
\widehat{P}_{\mathrm{LLL}}.
\label{eq:def H fs supplementary}
\end{equation}
Here, $\widehat{P}_{\mathrm{LLL}}$ is the projector onto the lowest Landau level (LLL), and the terms $\widehat{H}_{\mathrm{Hub}}$, $\widehat{H}_{\mathrm{Heis}}$, and $\widehat{H}_{\mathrm{trans}}$ are defined as follows.

\medskip
\noindent
The unprojected Hamiltonian in real space, in terms of local densities built from the four-flavor rotor spinor, reads:
\begin{subequations}\label{eq:unproj-H}
  \begin{align}
      \widehat{H} &= u\,\widehat{H}_{\mathrm{Hub}} - v\,\widehat{H}_{\mathrm{Heis}} + h\,\widehat{H}_{\mathrm{trans}}, \label{eq:unproj-H-a-supp} \\
      \widehat{H}_{\mathrm{Hub}} &= \sum_{a,b=1}^{N} \int_{S^2} d\Omega_a\, d\Omega_b\, U_{ab}\, \Bigl[n^{0}(\Omega_a)\, n^{0}(\Omega_b)\Bigr], \label{eq:unproj-H-b-supp} \\
      \widehat{H}_{\mathrm{Heis}} &= \sum_{a,b=1}^{N} \int_{S^2} d\Omega_a\, d\Omega_b\, U_{ab}\, \Bigl[n^{\mathcal R}(\Omega_a)\, n^{\mathcal R}(\Omega_b)\Bigr], \label{eq:unproj-H-c-supp} \\
      \widehat{H}_{\mathrm{trans}} &= \sum_{a=1}^{N} \int_{S^2} d\Omega_a\, n^{\mathcal L}(\Omega_a). \label{eq:unproj-H-d-supp}
  \end{align}
\end{subequations}
The local densities are defined by
\begin{align}
  n^{\alpha}(\Omega)
  = \begin{pmatrix}
    \hat\psi^{\dagger}_{0,0} & \hat\psi^{\dagger}_{1,-1} & \hat\psi^{\dagger}_{1,0} & \hat\psi^{\dagger}_{1,1}
  \end{pmatrix}
  \mathcal M^{\alpha}
  \begin{pmatrix}
    \hat\psi_{0,0} \\ \hat\psi_{1,-1} \\ \hat\psi_{1,0} \\ \hat\psi_{1,1}
  \end{pmatrix},
\end{align}
with $\mathcal M^0 = I$, $\mathcal M^{\mathcal L} = \mathrm{diag}(0,2,2,2)$, and $\mathcal M^{\mathcal R}$ the two-body rotor matrix elements (defined below via Eq.~\eqref{eq:R-matrix-def}).

\medskip
\noindent
The LLL projector acts on the field operator as
\begin{subequations}\label{eq:LLL-proj}
  \begin{align}
    \widehat{P}_{\mathrm{LLL}} \, \hat\psi(\Omega) \, \widehat{P}_{\mathrm{LLL}}
    &= \sum_{m=-Q}^{Q} Y_{Q,Q,m}(\Omega)\, \hat c_m, \label{eq:LLL-proj-a-supp} \\
    \mathbf{c}_m^{\dagger} &\equiv \bigl(c_{m,(0,0)}^{\dagger},\; c_{m,(1,-1)}^{\dagger},\; c_{m,(1,0)}^{\dagger},\; c_{m,(1,1)}^{\dagger}\bigr), \label{eq:LLL-proj-b-supp}
  \end{align}
\end{subequations}
where $Y_{Q,Q,m}$ are the LLL monopole harmonics and the four components of $\mathbf{c}_m^{\dagger}$ correspond to the truncated rotor flavors $\{(l,m)=(0,0),(1,-1),(1,0),(1,1)\}$. Here $m$ denotes the $z$-component of the angular momenta and it can take $2Q+1$ values from $-Q$ to $Q$.

\medskip
\noindent
Projecting the Hamiltonian explicitly to the LLL yields
\begin{subequations}\label{eq:H-LLL}
  \begin{align}
    \widehat{H}_{\mathrm{fzs}} &=
    u\,\widehat{H}_{\mathrm{Hub}}^{\mathrm{LLL}}
    - v\,\widehat{H}_{\mathrm{Heis}}^{\mathrm{LLL}}
    + h\,\widehat{H}_{\mathrm{trans}}^{\mathrm{LLL}}, \label{eq:H-LLL-a-supp} \\
    \widehat{H}_{\mathrm{Hub}}^{\mathrm{LLL}} &= \sum_{m_1,m_2,m=-Q}^{Q} V_{m_1, m_2, m_2-m, m_1+m}\; \mathbf{c}_{m_1}^{\dagger}\, \mathbf{c}_{m_2}^{\dagger}\, \mathbf{c}_{m_1+m}\, \mathbf{c}_{m_2-m}, \label{eq:H-LLL-b-supp} \\
    \widehat{H}_{\mathrm{Heis}}^{\mathrm{LLL}} &= \sum_{m_1,m_2,m=-Q}^{Q} V_{m_1, m_2, m_2-m, m_1+m}\; \mathbf{c}_{m_1}^{\dagger}\, \mathbf{c}_{m_2}^{\dagger}\, \mathcal{M}^{\mathcal{R}}\, \mathbf{c}_{m_1+m}\, \mathbf{c}_{m_2-m}, \label{eq:H-LLL-c-supp} \\
    \widehat{H}_{\mathrm{trans}}^{\mathrm{LLL}} &= \sum_{m=-Q}^{Q} \mathbf{c}_m^{\dagger}\, \mathcal{M}^{\mathcal{L}}\, \mathbf{c}_m. \label{eq:H-LLL-d-supp}
  \end{align}
\end{subequations}
Here $V_{m_1,m_2,m_3,m_4}$ are the usual LLL two-body matrix elements on the sphere. The rotor matrix elements $\mathcal{M}^{\mathcal{R}}$ acting on the four-flavor space is the truncated-basis matrix built from spherical-harmonic couplings; in components it reads
\begin{align}
\mathcal{M}^{\mathcal{R}}_{(l_1,m_1),(l_2,m_2);(l_3,m_3),(l_4,m_4)} &= \sum_{\alpha=-1,0,1} (-1)^{- m_1 - m_2 - \alpha}
\, \frac{\sqrt{9 (2 l_1 + 1) (2 l_2 + 1) (2 l_3 + 1) (2 l_4 + 1)}}{4 \pi} \nonumber \\
&\quad \times
\begin{pmatrix}
  l_1 & 1 & l_4 \\
  0 & 0 & 0
\end{pmatrix}
\begin{pmatrix}
  l_2 & 1 & l_3 \\
  0 & 0 & 0
\end{pmatrix}
\begin{pmatrix}
  l_1 & 1 & l_4 \\
  - m_1 & - \alpha & m_4
\end{pmatrix}
\begin{pmatrix}
  l_2 & 1 & l_3 \\
  - m_2 & \alpha & m_3
\end{pmatrix},
\label{eq:R-matrix-def}
\end{align}
which evaluates the interaction terms $\sum_{\alpha=-1,0,1} \hat{n}_i^{\alpha} \cdot \hat{n}_j^{\alpha}$ in the truncated $\{|l,m\rangle\}$ rotor basis with $l \in \{0,1\}$.

\medskip
\noindent
Eq.~\eqref{eq:H-LLL} provides the LLL-projected, second-quantized form used in our ED/DMRG calculations. The matrix $\mathcal{M}^{\mathcal{L}}$ encodes the single-site flavor weights, while $\mathcal{M}^{\mathcal{R}}$ captures the two-body rotor matrix elements within the truncated $\{(0,0),(1,-1),(1,0),(1,1)\}$ manifold. The coefficients $V_{m_1,m_2,m_3,m_4}$ contain interaction dependence through the standard monopole harmonic integrals on $\mathbb S^2$.

The interaction $U(\Omega_{ab})$ is encoded through matrix elements $V_{m_1, m_2, m_3, m_4}$ using Haldane pseudopotentials~\cite{Haldane_1983_FQHE}, following the convention established in Ref.~\cite{Zhu_Han_Huffman_Hofmann_He_2023}. These pseudopotentials provide a systematic decomposition of rotationally invariant interactions on the sphere into components of definite relative angular momenta.
These are expressed using Wigner-\(3j\) symbols as
\begin{align}
    V_{m_1, m_2, m_3, m_4} = \sum_{l=0}^{L} V_l (4s - 2l + 1)
    \begin{pmatrix}
        s & s & 2s-l \\
        m_1 & m_2 & -m_1 - m_2
    \end{pmatrix}
    \begin{pmatrix}
        s & s & 2s-l \\
        m_3 & m_4 & -m_3 - m_4
    \end{pmatrix}.
\end{align}
We restrict to nonzero \( V_0 \) and \( V_1 \). Pseudopotentials represent the projection of central interactions onto the lowest Landau level, decomposed by total angular momenta. Specifically, \( V_0 \) corresponds to the projection of the contact interaction \( \delta(\Bar{r} - \Bar{r}') \), while \( V_1 \) captures that of \( \nabla^2 \delta(\Bar{r} - \Bar{r}') \).
We determine the optimal values for the Hamiltonian parameters to be $V_0 = 6.5$, $V_1 = 1.0$, $u=1.0$, and $v = 1.4$, and all the simulations for the conformal data are done with these parameters.

\section{Truncated quantum rotor matrix elements}

\label{app: truncated rotor matrix elements}

We briefly recall the quantum rotor model: each site hosts a quantum rotor, whose configuration is described by a unit vector $\mathbf{n}$ on the sphere ($\mathbf{n}^2 = 1$). The components of $\mathbf{n}$ are the fundamental degrees of freedom, and the Hilbert space is spanned by spherical harmonics $|l, m\rangle$, as in the previous section.

The position operator $\hat{\mathbf{n}}$ acts as a vector operator in this space, with spherical components given by
\begin{align}
    n_q = \sqrt{\frac{4\pi}{3}}\, Y_{1,q}(\mathbf{\Omega}),
\end{align}
where $q = -1, 0, 1$ labels the  spherical basis, following the notation of the previous section.

To compute the matrix elements of the interaction term $\hat{\mathbf{n}}_i \cdot \hat{\mathbf{n}}_j$ in the truncated rotor basis, we evaluate
\begin{align}
    & \sum_{\alpha=-1,0,1} \bra{l_1, m_1} \bra{l_2, m_2} \hat{n}_i^{\alpha} \cdot \hat{n}_j^{\alpha} \ket{l_3, m_3} \ket{l_4, m_4} \nonumber \\
    & = \frac{4 \pi}{3}\sum_{\alpha=-1,0,1} \int d\mathbf{\Omega}_i\, d\mathbf{\Omega}_j\, Y^{*}_{l_1, m_1}(\mathbf{\Omega}_i) Y^{*}_{l_2, m_2}(\mathbf{\Omega}_j) \nonumber \\
    & \qquad \times Y^{*}_{1, \alpha}(\mathbf{\Omega}_i) Y_{1, \alpha}(\mathbf{\Omega}_j) Y_{l_3, m_3}(\mathbf{\Omega}_j) Y_{l_4, m_4}(\mathbf{\Omega}_i) \nonumber \\
    & = \frac{4 \pi}{3}\sum_{\alpha=-1,0,1} (-1)^{- m_1 - m_2 - \alpha} \frac{\sqrt{9 (2 l_1 + 1) (2 l_2 + 1) (2 l_3 + 1) (2 l_4 + 1)}}{4 \pi} \nonumber \\
    & \qquad \times
    \begin{pmatrix}
      l_1 & 1 & l_4 \\
      0 & 0 & 0
    \end{pmatrix}
    \begin{pmatrix}
      l_2 & 1 & l_3 \\
      0 & 0 & 0
    \end{pmatrix}
    \begin{pmatrix}
      l_1 & 1 & l_4 \\
      - m_1 & - \alpha & m_4
    \end{pmatrix}
    \begin{pmatrix}
      l_2 & 1 & l_3 \\
      - m_2 & \alpha & m_3
    \end{pmatrix}.
    \label{o3 rotor matrix elements}
\end{align}
This expression gives the explicit matrix elements of the two-site interaction in the truncated rotor basis, and corresponds to the matrix elements of $\frac{4 \pi}{3}\mathcal{M}^{\mathcal{R}}$ as shown in Eq.~\eqref{eq:R-matrix-def} in the previous section.

\section{Relation of truncated quantum rotor with bilayer heisenberg model}

Interestingly, the $O(3)$ rotor model has a correspondence with the bilayer Heisenberg antiferromagnet (AFM); in particular, the rotor can capture the low-energy physics of the bilayer AFM. In the strong inter-layer coupling limit, a bilayer AFM has, on every lattice site, a singlet ground state and a gapped triplet excitation formed by the two spins on that site.  These levels coincide with the \(l=0\) and \(l=1\) states of an \(O(3)\) quantum rotor once one identifies~\cite{Sachdev_2011}  
\begin{align}
\hat{\mathbf{L}}_i \;\longleftrightarrow\; \hat{\mathbf{S}}_{1,i} + \hat{\mathbf{S}}_{2,i},
\qquad
\hat{\mathbf{n}}_i \;\longleftrightarrow\; \hat{\mathbf{S}}_{1,i} - \hat{\mathbf{S}}_{2,i}.
\end{align}
so that the angular momenta of the rotor is the total spin of the rung and its orientation vector is the staggered (Néel) moment.  Intra-layer coupling of the antiferromagnet then hops the triplet between sites exactly as the rotor coupling \(-J\,\hat{\mathbf n}_i\!\cdot\!\hat{\mathbf n}_j\) propagates the \(l=1\) excitation, making the quantum rotor a compact, symmetry-preserving low-energy description of the bilayer magnet~\cite{Sachdev_2011}. 
Motivated by this correspondence, we keep only the singlet and triplet, i.e.\ set \(l_{\rm max}=1\), thereby retaining the full low-energy manifold while discarding higher-spin states that lie far above the gap.

\section{O($N$) generalization: Hamiltonian and matrix elements}

\noindent
We generalize to an $O(N)$ rotor with unit vector 
$\hat{\mathbf n}=(n^1,\dots,n^N)$ 
on $S^{N-1}$ and a truncated on-site Hilbert space spanned by hyperspherical harmonics 
$\{|l,\mu\rangle\}$, corresponding to rank-$l$ irreducible representations of $SO(N)$ (with $\mu$ denoting the degeneracy index within a given representation). 
In the $\{l=0\oplus 1\}$ truncation, the local ``flavor'' space is $(\mathbf{1}\oplus \mathbf{N})$, i.e.\ one singlet plus $N$ vector components. 
We denote the $(N+1)$-component field spinor as
\[
\hat\Psi^\dagger \equiv 
\bigl(
\hat\psi^\dagger_{0},\,
\hat\psi^\dagger_{1,1},\ldots,\hat\psi^\dagger_{1,N}
\bigr),
\]
where the first entry is the $l\!=\!0$ singlet and the rest are the $l\!=\!1$ vector components.

\medskip
\noindent
The $O(N)$ analogue of Eq.~\eqref{eq:def H fs supplementary} reads
\begin{equation}
\widehat{H}^{(N)}_{\mathrm{fzs}}
=
\widehat{P}_{\mathrm{LLL}}
\Bigl(
u\,\widehat{H}_{\mathrm{Hub}}
- v\,\widehat{H}^{(N)}_{\mathrm{Heis}}
+ h\,\widehat{H}_{\mathrm{trans}}
\Bigr)
\widehat{P}_{\mathrm{LLL}},
\label{eq:ON-H-def}
\end{equation}
with the unprojected real-space form
\begin{subequations}\label{eq:ON-unproj}
\begin{align}
\widehat{H} 
&= u\,\widehat{H}_{\mathrm{Hub}} 
- v\,\widehat{H}^{(N)}_{\mathrm{Heis}} 
+ h\,\widehat{H}_{\mathrm{trans}},\\
\widehat{H}_{\mathrm{Hub}}
&=\sum_{a,b=1}^{N_s}\!\int_{S^2}\!d\Omega_a\,d\Omega_b\,
U_{ab}\; n^{0}(\Omega_a)\,n^{0}(\Omega_b),\\
\widehat{H}^{(N)}_{\mathrm{Heis}}
&=\sum_{a,b=1}^{N_s}\!\int_{S^2}\!d\Omega_a\,d\Omega_b\,
U_{ab}\;\sum_{A=1}^{N} n^{A}(\Omega_a)\,n^{A}(\Omega_b),\\
\widehat{H}_{\mathrm{trans}}
&=\sum_{a=1}^{N_s}\int_{S^2}\!d\Omega_a\, n^{\mathcal L}(\Omega_a),
\end{align}
\end{subequations}
where $N_s$ is the number of electrons.
The local densities are bilinears 
$n^\alpha(\Omega)=\hat\Psi^\dagger(\Omega)\,\mathcal M^\alpha\,\hat\Psi(\Omega)$ with
\[
\mathcal M^{0}=\mathrm{diag}(1,1,\ldots,1),\qquad
\mathcal M^{\mathcal L}=\mathrm{diag}(0,w,\ldots,w),
\]
where $w$ fixes the relative weight of the $l\!=\!1$ sector (e.g.\ $w=2$ for $N=3$).

\medskip
\noindent
Projecting to the LLL yields
\begin{subequations}\label{eq:ON-LLL}
\begin{align}
\widehat{H}^{(N)}_{\mathrm{fzs}}
&=
u\,\widehat{H}_{\mathrm{Hub}}^{\mathrm{LLL}}
- v\,\widehat{H}^{(N)\!,\mathrm{LLL}}_{\mathrm{Heis}}
+ h\,\widehat{H}_{\mathrm{trans}}^{\mathrm{LLL}},\\
\widehat{H}_{\mathrm{Hub}}^{\mathrm{LLL}}
&=\!\!\sum_{m_1,m_2,m=-s}^{s}\!\! 
V_{m_1,m_2,m_2-m,m_1+m}\;
\mathbf c^\dagger_{m_1}\,\mathbf c^\dagger_{m_2}\,
\mathbf c_{m_1+m}\,\mathbf c_{m_2-m},\\
\widehat{H}^{(N)\!,\mathrm{LLL}}_{\mathrm{Heis}}
&=\!\!\sum_{m_1,m_2,m=-s}^{s}\!\!
V_{m_1,m_2,m_2-m,m_1+m}\;
\mathbf c^\dagger_{m_1}\,\mathbf c^\dagger_{m_2}\,
\mathcal M^{\mathcal R,(N)}\,
\mathbf c_{m_1+m}\,\mathbf c_{m_2-m},\\
\widehat{H}_{\mathrm{trans}}^{\mathrm{LLL}}
&=\sum_{m=-s}^{s}\mathbf c_m^\dagger\,
\mathcal M^{\mathcal L}\,\mathbf c_m,
\end{align}
\end{subequations}
where $\mathbf c_m^\dagger$ creates the $(\mathbf{1}\oplus\mathbf N)$ flavor at orbital $m$, 
and $V_{m_1,m_2,m_3,m_4}$ are the standard LLL two-body matrix elements (constructed via pseudopotentials). 
The $N$-dependence enters only through $\mathcal M^{\mathcal R,(N)}$.

\paragraph{Quantum numbers and degeneracies.}
For an $O(N)$ rotor, the wavefunctions are hyperspherical harmonics 
$Y^{(N)}_{l\mu}(\Omega)$ on $S^{N-1}$, which transform as the symmetric traceless rank-$l$ representations of $SO(N)$. 
They are labeled by a single angular momenta quantum number $l=0,1,2,\dots$ 
and a multi-index $\mu$ that distinguishes the degenerate components within a given $l$.
The degeneracy of the level $l$ is
\begin{equation}
d_l = \frac{(2l + N - 2)(l + N - 3)!}{l!\,(N - 2)!},
\end{equation}
so that $d_0 = 1$ (singlet) and $d_1 = N$ (vector representation).
Thus, the $\{l=0,1\}$ truncation yields a local Hilbert space of dimension $1+N$.

\paragraph{Two-body rotor matrix elements.}
The two-body matrix elements that enter the $O(N)$ rotor Hamiltonian, denoted $\mathcal M^{\mathcal R,(N)}$, encode the internal structure of the model and depend on the overlaps between single-particle states and the components of the unit vector $\hat{\mathbf n}$ on the sphere. While their explicit analytic form involves group-theoretic quantities such as Clebsch–Gordan coefficients and integrals over hyperspherical harmonics, in practice these matrix elements can be efficiently computed numerically for any $N$ and for the truncated $(\mathbf{1}\oplus\mathbf N)$ Hilbert space used in our simulations.

\medskip
\noindent
With this approach, Eq.~\eqref{eq:ON-LLL} defines the $O(N)$ extension of the truncated quantum rotor Hamiltonian: the monopole harmonic coefficients $V_{m_1,m_2,m_3,m_4}$ remain unchanged, while the internal structure enters only via $\mathcal M^{\mathcal L}$ and $\mathcal M^{\mathcal R,(N)}$, enabling straightforward ED/DMRG simulations with an $O(N)$ rotor flavor space.

\section{Conformal primaries and descendants}

\noindent
In conformal field theory, operators are organized into families. Each family consists of a primary operator and its descendants, which are generated by acting with momentum operators (derivatives).

A primary operator $\mathcal{O}$ is defined by its transformation properties under conformal symmetry. At the origin (in radial quantization), it satisfies:
\begin{align}
  [K_i, \mathcal O(0)] = 0,\qquad D\,\mathcal O(0) = \Delta_{\mathcal O}\, \mathcal O(0),
\end{align}
where $K_i$ are special conformal generators, $D$ is the dilatation operator, and $\Delta_{\mathcal O}$ is the scaling dimension.

Descendants are generated by applying momentum operators (derivatives) to the primary operator:
\begin{subequations}
\begin{align}
  \text{Level 1: } &\quad \partial_\mu \mathcal{O} \\
  \text{Level 2: } &\quad \partial_\mu \partial_\nu \mathcal{O} \\
  \text{Level $n$: } &\quad \partial_{\mu_1} \partial_{\mu_2} \cdots \partial_{\mu_n} \mathcal{O}
\end{align}
\end{subequations}

On $\mathbb{R} \times \mathbb{S}^2$, this translates to simple energy rules: the primary creates a state with energy $\Delta_{\mathcal O}$, and each application of a momentum operator $P_i$ increases the energy by exactly 1. Therefore, all level-$n$ descendants have energies $\Delta_{\mathcal O} + n$. Throughout the following discussion, we use the notation $L_P$ to denote the angular momenta of the primary operator and $L_D$ to denote the angular momenta of its descendants.

\subsection{Angular momenta structure of descendants}

\noindent
For scalar primaries i.e. primaries with angular momenta $L_P = 0$, at descendant level $n$, the allowed angular momenta are:
\begin{align}
  L_D \in \{n, n-2, n-4, \ldots\} \quad \text{down to} \quad
  \begin{cases}
  0, & \text{if } n \text{ is even} \\
  1, & \text{if } n \text{ is odd}
  \end{cases}
\end{align}

\noindent
For spinning primaries i.e. primaries with $L_P > 0$, at descendant level $n$, the allowed angular momenta are:
\begin{align}
  L_D \in \{L_P + n, L_P + n - 2, L_P + n - 4, \ldots, |L_P - n|\}
\end{align}

\noindent
Each angular momenta multiplet $L$ contains $2L + 1$ states with magnetic quantum numbers $m = -L, -L+1, \ldots, +L-1, +L$.

\subsection{Examples}

\noindent\textbf{Scalar primary ($L_P = 0$):}
\begin{itemize}
  \item Level $n = 1$: Energy $\Delta + 1$, angular momenta $L_D = 1$ (corresponds to $\partial_\mu \mathcal{O}$)
  \item Level $n = 2$: Energy $\Delta + 2$, angular momenta $L_D = 2, 0$ (corresponds to $\partial_\mu \partial_\nu \mathcal{O}$)
  \item Level $n = 3$: Energy $\Delta + 3$, angular momenta $L_D = 3, 1$ (corresponds to $\partial_\mu \partial_\nu \partial_\rho \mathcal{O}$)
\end{itemize}

\noindent\textbf{Vector primary ($L_P = 1$):}
\begin{itemize}
  \item Level $n = 1$: Energy $\Delta + 1$, angular momenta $L = 2, 1, 0$
  \begin{itemize}
    \item $L = 2$: corresponds to traceless symmetric part of $\partial_\mu \mathcal{O}_\nu$
    \item $L = 1$: corresponds to antisymmetric part $\partial_{[\mu} \mathcal{O}_{\nu]}$
    \item $L = 0$: corresponds to trace part $\partial_\mu \mathcal{O}^\mu$
  \end{itemize}
  \item Level $n = 2$: Energy $\Delta + 2$, angular momenta $L = 3, 2, 1, 0$ (corresponds to $\partial_\mu \partial_\rho \mathcal{O}_\nu$)
  % \item The antisymmetric combination $\epsilon_{\mu\nu\rho} \partial_\mu \mathcal{O}_\nu$ creates a descendant at level $n = 1$ with angular momenta $L = 2, 0$ (the $L = 1$ component is eliminated by the antisymmetrization)
\end{itemize}

\subsection{Conserved quantities and their descendants}

\noindent
Conserved quantities are somewhat special since they are generators of symmetries, as a consequence they have a very special descendant structure. The most important examples are:

\vspace{1em}
\noindent
\noindent\textbf{Stress-energy tensor ($T_{\mu\nu}$):}
\noindent The stress-energy tensor is a symmetric, traceless, conserved tensor that generates spacetime symmetries (translations, rotations, scale transformations, and special conformal transformations). On $\mathbb{R} \times \mathbb{S}^2$:
\begin{itemize}
  \item Primary: $T_{\mu\nu}$ with scaling dimension $\Delta = 3$ and angular momenta $L_P = 2$
  \item Level $n = 1$ descendants: Energy $\Delta + 1 = 4$, angular momenta $L_D = 3, 2, 1$
  \begin{itemize}
    \item $L = 3$: traceless symmetric part of $\partial_\rho T_{\mu\nu}$
    \item $L = 2$: mixed symmetry combinations
    \item $L = 1$: trace and divergence parts $\partial_\rho T^{\rho\nu} = 0$ (vanishes due to conservation)
  \end{itemize}
\end{itemize}

\noindent
\noindent\textbf{Noether current ($J_\mu$):}
\noindent For internal symmetries like $O(3)$, the Noether current is a conserved vector that generates internal rotations:
\begin{itemize}
  \item Primary: $J_\mu$ with scaling dimension $\Delta = 2$ and angular momenta $L_P = 1$
  \item Level $n = 1$ descendants: Energy $\Delta + 1 = 3$, angular momenta $L = 2, 1, 0$
  \begin{itemize}
    \item $L = 2$: traceless symmetric part of $\partial_\rho J_\mu$
    \item $L = 1$: antisymmetric part $\partial_{[\rho} J_{\mu]}$
    \item $L = 0$: divergence $\partial_\mu J^\mu = 0$ (vanishes due to conservation)
  \end{itemize}
\end{itemize}

\noindent
The conservation laws $\partial_\mu T^{\mu\nu} = 0$ and $\partial_\mu J^\mu = 0$ eliminate certain descendant states, leading to shortened multiplets compared to generic operators.

\section{Conformal perturbation theory}

For any $D$-dimensional CFT on $\mathbb{R} \times
\mathbb{S}^{D-1}$, the state-operator correspondence
allows to reinterpret the  dilatation operator as a
``quantum Hamiltonian''
$\widehat{H}_{\mathrm{CFT}}$
whose eigenstate $\ket{\mathrm{o}}$
has as eigenvalue the scaling dimension $\Delta_{\mathrm{o}}$
of a CFT operator labeled by $\mathrm{o}$, 
\begin{align}
\widehat{H}_{\mathrm{CFT}}\,\ket{\mathrm{o}}=
\Delta_{\mathrm{o}}\,\ket{\mathrm{o}}\red{.}
\label{appeq:def HCFT}
\end{align}

Conformal perturbation theory provides a systematic framework to describe
deviations from criticality
originating either from a CFT-breaking perturbation
or from finite-size effects.
Following Ref.~\cite{Icosahedron,Ising_CPT},
the effective Hamiltonian
in Eq.\ (\ref{appeq:def HCFT})
is replaced by $\widehat{H}:=\widehat{H}_{\mathrm{CFT}}+\delta\widehat{H}$,
where the perturbation is expressed as
\begin{align}
\delta\widehat{H} =
\sum_n
\int_{\mathbb{S}^{D-1}}d\mathbf{\Omega}\,
g_n(\mathbf{\Omega})\,
\widehat{\chi}_n(0,\mathbf{\Omega}),
\end{align}
with $\widehat{\chi}_n$ local CFT operators inserted at $\tau = 0$ on
$\mathbb{R}\times\mathbb{S}^{D-1}$, and $g_n(\mathbf{\Omega})$ the
corresponding coupling functions. The perturbing operators must
respect the symmetries of the microscopic theory. \red{F}or the O(3) rotor
model, $\widehat{\chi}_n$
must preserve both SO(3) rotational symmetry and O(3)
internal spin symmetry.

The energy eigenvalues of $H$ are then given by
\begin{align}
E_{\mathrm{o}_m}=&\,
E_{\mathrm{o}_m, \mathrm{CFT}}
+
\bra{\mathrm{o}_m} \delta\widehat{H}\ket{\mathrm{o}_m}
\nonumber\\
&\,
+
\sum_{i \neq m}
\frac{
\bra{\mathrm{o}_m}
\delta \widehat{H}
\ket{\mathrm{o}_i}
\bra{\mathrm{o}_i}
\delta \widehat{H}
\ket{\mathrm{o}_m}
     }
     {
E_{\mathrm{o}_m,\mathrm{CFT}}
-
E_{\mathrm{o}_i,\mathrm{CFT}}
     }
\nonumber\\
&\,
+
\mathcal{O}\left((\delta \widehat{H})^3\right),
\end{align}
where
$E_{\mathrm{o}_m, \mathrm{CFT}}\equiv\Delta_{\mathrm{o}_m}$.
To first order, only the diagonal matrix elements
$\bra{\mathrm{o}_m} \delta \widehat{H} \ket{\mathrm{o}_m}$ contribute.
These are determined by OPE coefficients between
$\widehat{\chi}_n$ and $\widehat{\mathrm{o}}_m$.
For $\mathrm{o}_m$ labeling a scalar primary
and $\partial \mathrm{o}_m$ labeling its first descendant,
the first-order shifts are
\begin{subequations}
\begin{align}
\delta E_{\mathrm{o}_m} &=
  g_{\chi_n}\,
  f_{\mathrm{o}_m, \chi_n, \mathrm{o}_m},
  \label{eq:deltaE_om} \\[4pt]
\delta E_{\partial \mathrm{o}_m} &=
  g_{\chi_n}\,
  f_{\mathrm{o}_m, \chi_n, \mathrm{o}_m}\,
  \mathcal{A}_{\mathrm{o}_m,\chi_n}.
  \label{eq:deltaE_partial_om}
\end{align}
\end{subequations}
respectively. Here,
$g_{\chi_n} = \int_{S^{D-1}} g_n(\mathbf{\Omega})$
is the integrated coupling,
while the multiplicative factor $\mathcal{A}_{\mathrm{o}_m,\chi_n}$ is
fixed by the conformal symmetry to be
\begin{align}
    \mathcal{A}_{\mathrm{o}_m,\chi_n} = 1 + \frac{\Delta_{\chi_n} (\Delta_{\chi_n} - 3)}{6 \Delta_{\mathrm{o}_m}}.
\end{align}

\section{Large-$S$ expansion}

If the CFT has a continuous internal symmetry group $\mathrm{G}$, there exists a conserved Noether current and a corresponding conserved charge operator $\widehat{Q}$, with eigenvalues $Q$ that scale with the area $R^{D-1}$ in the thermodynamic limit. We define the charge and energy densities as
\begin{subequations}
\begin{align}
\rho &:= \frac{Q}{R^{D-1}}, \label{eq:rho_def} \\
\delta\varepsilon &:= \frac{\delta E}{R^{D-1}}. \label{eq:energy_density_def}
\end{align}
\end{subequations}
To study states at fixed $\rho$ and $\delta\varepsilon$, we introduce a chemical potential $\mu$ and consider the Hamiltonian $\widehat{H}_{\mathrm{CFT}} + \mu \widehat{Q}$. Dimensional analysis then gives
\begin{subequations}
\begin{align}
\rho &= c_\rho\, \mu^{D-1}, \label{eq:rho_mu} \\
\delta\varepsilon &= c_\varepsilon\, \mu^D, \label{eq:energy_mu}
\end{align}
\end{subequations}
where $c_\rho$ and $c_\varepsilon$ are dimensionless constants of order one. Inverting, we have
\begin{subequations}
\begin{align}
\mu &= \left( \frac{\rho}{c_\rho} \right)^{1/(D-1)}, \label{eq:mu_rho} \\
\delta\varepsilon &= c\, \rho^{D/(D-1)}, \qquad c = c_\varepsilon\, c_\rho^{-D/(D-1)}. \label{eq:energy_rho}
\end{align}
\end{subequations}
Combining state-operator correspondence with Eqs.~\eqref{eq:energy_density_def}, and \eqref{eq:energy_rho}, we obtain the leading scaling of the scaling dimension at large charge:
\begin{equation}
\Delta_{o} = c\, Q^{D/(D-1)}.
\label{eq:largeQ_leading}
\end{equation}

This leading behavior is the first term in a systematic large-charge (semiclassical) expansion~\cite{Hellerman15,Monin17,Cuomo2020,Cuomo2021}:
\begin{equation}
\Delta^{\mathrm{sc}}_{\mathcal{O}} = \sum_{n=0}^\infty c^{\mathrm{sc}}_n\, Q^{\frac{D}{D-1} - n},
\label{eq:largeQ_semiclassical}
\end{equation}
where the coefficients $c^{\mathrm{sc}}_n$ can be computed in effective field theory. The expansion in Eq.~\eqref{eq:largeQ_semiclassical} does not include a $Q^0$ term at leading order, but such a term appears at one-loop order~\cite{Monin17}:
\begin{equation}
\Delta^{\mathrm{sc+1-loop}}_{o} = \Delta^{\mathrm{sc}}_{o} + c^{\mathrm{1-loop}}.
\label{eq:largeQ_1loop}
\end{equation}
Explicit methods for computing these coefficients are available in the literature~\cite{Hellerman15,Alvarez-Gaume17,Monin17,Banerjee18,Banerjee19,Alvarez-Gaume19,Watanabe21,Giombi21,Gaume21,Cuomo2020,Cuomo2021}. In what follows, we focus on the case $D=3$ and $\mathrm{G} = \mathrm{O}(3)$.

In the thermodynamic limit $R \propto \sqrt{N} \to \infty$ at zero temperature, the phase diagram of the fuzzy sphere Hamiltonian $\widehat{H}_{\mathrm{fzs}}(R)$ consists of: (i) a long-range ordered phase that spontaneously breaks SO(3) down to SO(2); (ii) a quantum disordered phase with a gapped, non-degenerate, symmetric ground state; and (iii) a quantum critical point separating the two, described by the $(2+1)$-dimensional O(3) Wilson-Fisher CFT. In the ordered phase, Goldstone's theorem predicts two gapless modes (transverse to the order parameter), while the longitudinal mode (associated with the unbroken SO(2)) is not sharp. Approaching the critical point from the ordered side, the longitudinal gap closes continuously. In the disordered phase, all modes are gapped and strongly interacting.

If we instead consider the Hamiltonian $\widehat{H}_{\mathrm{fzs}}(R) + \mu \widehat{Q}^z$, where $\widehat{Q}^z$ is the conserved charge associated with the Cartan subalgebra of $\mathfrak{so}(3)$, the situation changes: the chemical potential $\mu$ explicitly breaks time-reversal symmetry and precludes emergent Lorentz invariance~\cite{Lange65,Nielsen76,Watanabe11,Watanabe12}. Of the two Goldstone modes present at $\mu=0$, only one remains gapless (the phonon), while the other becomes a sharp but gapped (pseudo-)Goldstone mode, with a gap set by $\mu$~\cite{Nicolis13,Watanabe13}. The scale $\mu$ introduces a UV length scale $1/\mu \propto R/\sqrt{S}$, where $S \gg 1$ is the eigenvalue of $\widehat{Q}^z$ for the states of interest. In this regime, the CFT at large charge is described by a weakly interacting effective theory for the gapless and gapped Goldstone modes, and the expansion~\eqref{eq:largeQ_1loop} applies.

Let us now make the large-$S$ expansion explicit for the lowest-lying states in each symmetry sector. For the scalar primary with $S^{\kappa(S)} = 1^-, 2^+, \ldots$ (with $\kappa(S) = (-1)^S$) and $L=0$, the scaling dimension admits the expansion
\begin{equation}
\begin{split}
\Delta_0(S^{\kappa(S)}, L=0) =\;&
\alpha\, (S^{\kappa(S)})^{3/2}
+ \beta\, (S^{\kappa(S)})^{1/2}
\\
& - 0.0937256
+ \gamma\, (S^{\kappa(S)})^{-1/2}
\\
& + \mathcal{O}\left(1/S^{\kappa(S)}\right),
\end{split}
\label{eq:largeS_scalar}
\end{equation}
where $\alpha$, $\beta$, and $\gamma$ are Wilson coefficients. These can be fixed by matching to the scaling dimensions of the lowest scalar primaries, as estimated from the conformal bootstrap (CB):
\begin{equation}
\begin{split}
\Delta_\sigma(1^-, 0) &\approx 0.518936, \\
\Delta_{t_2}(2^+, 0) &\approx 1.20954, \\
\Delta_{t_4}(4^+, 0) &\approx 2.99056.
\end{split}
\end{equation}
Solving, we find
\begin{equation}
\alpha \approx 0.31076, \qquad
\beta \approx 0.29818, \qquad
\gamma \approx 0.00370.
\label{eq:largeS_coeffs}
\end{equation}
This expansion captures the scaling of the lowest-energy state in each $S^{\kappa(S)}$ sector, and the alternation of parity $\kappa(S) = (-1)^S$ is confirmed numerically.

For the lowest-lying state in the $S^{\kappa(S+1)} = 1^+, 2^-, \ldots$ sector (with $\kappa(S) = (-1)^S$) and $L=0,1,\ldots$, corresponding to the gapped Goldstone mode, the scaling dimension is predicted~\cite{Cuomo2020,Cuomo2021} to be
\begin{subequations}
\label{eq:largeS_gappedGoldstone}
\begin{equation}
\begin{split}
\Delta(S^{\kappa(S+1)}, L) =\;&
\Delta_0(S^{\kappa(S)}, 0)
+ \mu(S^{\kappa(S)})
\\
& + c\, \frac{L(L+1)}{2\mu(S^{\kappa(S)})}
+ \mathcal{O}\left(\frac{L^4}{\mu^3(S^{\kappa(S)})}\right),
\end{split}
\label{eq:largeS_gappedGoldstone_a}
\end{equation}
where
\begin{equation}
\mu(S) := \frac{\partial \Delta_0(S, 0)}{\partial S}
\label{eq:largeS_gappedGoldstone_b}
\end{equation}
\end{subequations}
is the chemical potential, and $c$ is a Wilson coefficient. The term proportional to $L(L+1)$ describes the gapped dispersion of the pseudo-Goldstone mode. The value of $c$ can be fixed by matching to the scaling dimension of the conserved current operator, $\Delta_{j^\mu}(1^+, 1) = 2$.

The gapless phonon mode is also captured in the large-$S$ expansion. For the $S^{\kappa(S)} = 1^-, 2^+, \ldots$ sector with $L=0,1,2,\ldots$, the scaling dimension is given by~\cite{Cuomo2020,Cuomo2021}
\begin{equation}
\begin{split}
\Delta_0(S^{\kappa(S)}, L) =\;&
\Delta_0(S^{\kappa(S)}, 0)
+ \sum_L n_L\, \omega_L,
\\
\omega_L &= \sqrt{L(L+1)/2},
\end{split}
\label{eq:largeS_phonon}
\end{equation}
where $n_L$ is the occupation number of the phonon with angular momenta $L$ and single-particle energy $\omega_L$. Phonons with $L=1$ generate descendants, while $L\geq 2$ correspond to new primary operators.

\section{Details on the simulations}
We use exact diagonalization (ED) and density matrix renormalization group (DMRG) to determine the critical point and extract the conformal data. In ED, we exploit all available symmetries of the Hamiltonian to reduce computational cost and access larger system sizes.

\paragraph*{ED simulations.---}
We first implement the $U(1)$ symmetry associated with orbital $SO(3)$ by conserving $L_z$, which partitions the Hilbert space into sectors labeled by $L_z$. Although ED does not resolve the total angular momenta $L$, each $L$ multiplet spans eigenstates with $L_z = -L, \dots, L$. This structure can be inverted to recover $L$ by comparing eigenvalues across different $L_z$ sectors. For example, if an eigenvalue is present at $L_z = 3$ but absent at $L_z = 4$, it must belong to an $L = 3$ multiplet. By recursively eliminating contributions from higher-$L$ multiplets in lower-$L_z$ sectors, the remaining unmatched eigenvalues at $L_z = \ell$ can be uniquely assigned to $L = \ell$. This method enables multiplet resolution even when only $L_z$ is conserved.

Further refinement is achieved by exploiting the $L_z \to -L_z$ symmetry in the $L_z = 0$ sector. This symmetry allows the Hilbert space at $L_z = 0$ to be decomposed into even and odd parity sectors, depending on the behavior of the eigenstate under $L_z$ sign reversal. Since even-$L$ multiplets have even parity under this transformation and odd-$L$ multiplets have odd parity, the $L_z=0$ even and odd sectors effectively separate even-$L$ and odd-$L$ contributions. For instance, an eigenvalue identified as part of an $L=3$ multiplet from the $L_z=3$ sector must have a corresponding $L_z=0$ component in the odd-parity sector. Once such known higher-$L$ contributions are removed from the $L_z=0$ odd sector, any remaining eigenvalues must arise from the $L=1$ multiplet. Consequently, the $L_z=1$ sector becomes redundant for multiplet reconstruction, as all information it contains can be extracted from symmetry-resolved $L_z=0$ data.

Similarly, we use the $U(1)$ symmetry generated by $S_z$ to resolve spin sectors, along with the symmetry $S_z \to -S_z$ in the $S_z = 0$ sector. This splits the $S_z = 0$ space into even and odd parity sectors and allows us to reconstruct spin multiplets in the same manner as for angular momenta.

We also utilize the discrete $Z_2$ subgroup of the internal $O(3)$ symmetry. This symmetry classifies states according to their parity under internal inversion and further reduces the Hilbert space by separating sectors that do not mix.

Combining all these symmetries, the largest Hilbert space sector becomes $(L_z, S_z) = (2,2)$ instead of the typical $(0,0)$. Using this fully symmetry-resolved basis, we simulate systems with up to 12 electrons in 48 orbitals. The full Hilbert space of dimension around $70$ billion is reduced to around $300$ million in the largest sector. ED is parallelized across all symmetry sectors.

\paragraph*{DMRG simulations.---}
DMRG simulations use the same base code as Ref. \onlinecite{Ising_CPT}.
We use \href{https://github.com/ITensor/ITensors.jl}{ITensors.jl} \cite{ITensor, ITensor-r0.3} as the underlining tensor network library.
The four degrees of freedom representing the $O(3)$ rotor on the fuzzy sphere are represented as a chain of $4 \times N$ spinless electrons, in ascending order of orbital momenta.
We implemented an Abelian subgroup of O(3) (conservation of $l_z$ and of the parity of $l$), as well as the U(1) conservation of the electronic charge and the orbital momenta.
The MPO were computed as standard, discarding contributions of amplitude $V_{m_1, m_2, m_3, m_4}$ below $10^{-12}$ and compressed using the default compression of ITensors.jl.
We use noisy two-site DMRG \cite{White1992, White2005} to be ergodic in phase space due to the complexity of the model.
The amplitude of the noise was generally kept at $10^{-5}$.
Our convergence criterion at a given $\chi$ are variations of the energy and the mid-chain entropy below $10^{-7}$.
To compute excited states, we successively incorporate into the effective Hamiltonian the weighted projectors onto previously-computed low-energy states.

% \section{Mapping between truncated rotors and spin models}
% For $O(2)$ truncated rotor model, the

% \section{Free $O(3)$ CFT}

% \noindent
% The free boson on a sphere given by solution of the Laplacian on the surface of the sphere that is
% \begin{align}
%     \nabla^2 \phi = 0.
%     \label{eq: free boson}
% \end{align}
% The solution of this equation is given by the spherical harmonics $Y_{l, m} (\theta, \phi)$, and the eigenvalues are $E_{l, m} = l + \frac{1}{2}$, where $l \in \mathbb{Z}^{+} \cup 0$. For bosonic particles, you can now fill the energy levels to obtain the Fock space, which is resolved in $L_z = \sum_{\text{all particles}} m$. Then by multiplet resolving, one can obtain the

\section{Scaling dimension data for all symmetry sectors}

\subsection{$S=0:$}
\insertfigure{0}{+}
\insertfigure{0}{-}
\inserttable{0}{+}{0}
\inserttable{0}{+}{1}
\inserttable{0}{+}{2}
\inserttable{0}{+}{3}
\inserttable{0}{+}{4}
\inserttable{0}{+}{5}
\inserttable{0}{+}{6}
\inserttable{0}{+}{7}

\inserttable{0}{-}{0}
\inserttable{0}{-}{1}
\inserttable{0}{-}{2}
\inserttable{0}{-}{3}
\inserttable{0}{-}{4}
\inserttable{0}{-}{5}
\inserttable{0}{-}{6}
\inserttable{0}{-}{7}

\FloatBarrier
\newpage

\subsection{$S=1:$}
\insertfigure{1}{+}
\insertfigure{1}{-}
\inserttable{1}{+}{0}
\inserttable{1}{+}{1}
\inserttable{1}{+}{2}
\inserttable{1}{+}{3}
\inserttable{1}{+}{4}
\inserttable{1}{+}{5}
\inserttable{1}{+}{6}
\inserttable{1}{+}{7}
\inserttable{1}{-}{0}
\inserttable{1}{-}{1}
\inserttable{1}{-}{2}
\inserttable{1}{-}{3}
\inserttable{1}{-}{4}
\inserttable{1}{-}{5}
\inserttable{1}{-}{6}
\inserttable{1}{-}{7}
\FloatBarrier
\newpage

\subsection{$S=2:$}
\insertfigure{2}{+}
\insertfigure{2}{-}
\inserttable{2}{+}{0}
\inserttable{2}{+}{1}
\inserttable{2}{+}{2}
\inserttable{2}{+}{3}
\inserttable{2}{+}{4}
\inserttable{2}{+}{5}
\inserttable{2}{+}{6}
\inserttable{2}{+}{7}
\inserttable{2}{-}{0}
\inserttable{2}{-}{1}
\inserttable{2}{-}{2}
\inserttable{2}{-}{3}
\inserttable{2}{-}{4}
\inserttable{2}{-}{5}
\inserttable{2}{-}{6}
\inserttable{2}{-}{7}
\FloatBarrier
\newpage

\subsection{$S=3:$}
\insertfigure{3}{+}
\insertfigure{3}{-}
\inserttable{3}{+}{0}
\inserttable{3}{+}{1}
\inserttable{3}{+}{2}
\inserttable{3}{+}{3}
\inserttable{3}{+}{4}
\inserttable{3}{+}{5}
\inserttable{3}{+}{6}
\inserttable{3}{+}{7}
\inserttable{3}{-}{0}
\inserttable{3}{-}{1}
\inserttable{3}{-}{2}
\inserttable{3}{-}{3}
\inserttable{3}{-}{4}
\inserttable{3}{-}{5}
\inserttable{3}{-}{6}
\inserttable{3}{-}{7}
\FloatBarrier
\newpage

\subsection{$S=4:$}
\insertfigure{4}{+}
\insertfigure{4}{-}
\inserttable{4}{+}{0}
\inserttable{4}{+}{1}
\inserttable{4}{+}{2}
\inserttable{4}{+}{3}
\inserttable{4}{+}{4}
\inserttable{4}{+}{5}
\inserttable{4}{+}{6}
\inserttable{4}{+}{7}
\inserttable{4}{-}{0}
\inserttable{4}{-}{1}
\inserttable{4}{-}{2}
\inserttable{4}{-}{3}
\inserttable{4}{-}{4}
\inserttable{4}{-}{5}
\inserttable{4}{-}{6}
\inserttable{4}{-}{7}
\FloatBarrier
\newpage

\subsection{$S=5:$}
\insertfigure{5}{+}
\insertfigure{5}{-}
\inserttable{5}{+}{0}
\inserttable{5}{+}{1}
\inserttable{5}{+}{2}
\inserttable{5}{+}{3}
\inserttable{5}{+}{4}
\inserttable{5}{+}{5}
\inserttable{5}{+}{6}
\inserttable{5}{+}{7}
\inserttable{5}{-}{0}
\inserttable{5}{-}{1}
\inserttable{5}{-}{2}
\inserttable{5}{-}{3}
\inserttable{5}{-}{4}
\inserttable{5}{-}{5}
\inserttable{5}{-}{6}
\inserttable{5}{-}{7}
\FloatBarrier
\newpage

\subsection{$S=6:$}
\insertfigure{6}{+}
\insertfigure{6}{-}
\inserttable{6}{+}{0}
\inserttable{6}{+}{1}
\inserttable{6}{+}{2}
\inserttable{6}{+}{3}
\inserttable{6}{+}{4}
\inserttable{6}{+}{5}
\inserttable{6}{+}{6}
\inserttable{6}{+}{7}
\inserttable{6}{-}{0}
\inserttable{6}{-}{1}
\inserttable{6}{-}{2}
\inserttable{6}{-}{3}
\inserttable{6}{-}{4}
\inserttable{6}{-}{5}
\inserttable{6}{-}{6}
\inserttable{6}{-}{7}
\FloatBarrier
\newpage

\subsection{$S=7:$}
\insertfigure{7}{+}
\insertfigure{7}{-}
\inserttable{7}{+}{0}
\inserttable{7}{+}{1}
\inserttable{7}{+}{2}
\inserttable{7}{+}{3}
\inserttable{7}{+}{4}
\inserttable{7}{+}{5}
\inserttable{7}{+}{6}
\inserttable{7}{+}{7}
\inserttable{7}{-}{0}
\inserttable{7}{-}{1}
\inserttable{7}{-}{2}
\inserttable{7}{-}{3}
\inserttable{7}{-}{4}
\inserttable{7}{-}{5}
\inserttable{7}{-}{6}
\inserttable{7}{-}{7}
\FloatBarrier
\newpage

\section{OPE Data for all sectors}
% We present data and plots for selected OPE coefficients of the $O(3)$ WF CFT in Fig.~\ref{fig:ope-all}. Solid green lines denote reference values from conformal bootstrap, while dashed green lines correspond to those obtained by rescaling with descendant factors from CPT~\cite{Icosahedron}.

\subsection{$S=0:$}
\opefigure{0}
\opetable{0}{+}
\opetable{0}{-}
\FloatBarrier
\newpage

\subsection{$S=1:$}
\opefigure{1}
\opetable{1}{+}
\opetable{1}{-}
\FloatBarrier
\newpage

\subsection{$S=2:$}
\opefigure{2}
\opetable{2}{+}
\opetable{2}{-}
\FloatBarrier
\newpage

\subsection{$S=3:$}
\opefigure{3}
\opetable{3}{+}
\opetable{3}{-}
\FloatBarrier

\newpage
\subsection{$S=4:$}
\opetable{4}{+}
\opetable{4}{-}
\FloatBarrier

\subsection{$S=5:$}
\opetable{5}{-}
\FloatBarrier

\subsection{$S=6:$}
\opetable{6}{+}
\FloatBarrier

\bibliographystyle{apsrev4-2}
\bibliography{bibtex}